\documentclass[aps,prl,reprint,superscriptaddress,super,superbib,amsmath,showkeys,notitlepage]{revtex4-1}
\usepackage{color}
\usepackage[pdftex]{graphicx}  
\usepackage{amssymb}

\newcommand{\ew}[1]{\pmb{\big\langle} #1 \pmb{\big\rangle}}
\newcommand{\ket}[1]{| #1 \rangle}

\newcommand {\hc}{\text{H.c.}}

\newcommand{\lk}{\left(}
\newcommand{\rk}{\right)}
\newcommand{\lsz}{\left[}
\newcommand{\rsz}{\right]}
\newcommand{\lka}{\left\{}
\newcommand{\rka}{\right\}}
\usepackage[dvipsnames]{xcolor} %Allows me to highlight/color text

\newcommand{\Hop}{\hat{H}}
\newcommand{\Uop}{\hat{U}}

\newcommand{\aop}{\hat{a}}
\newcommand{\adop}{\hat{a}^\dagger}
\newcommand{\bop}{\hat{b}}
\newcommand{\bdop}{\hat{b}^\dagger}

\newcommand{\sigmop}{\hat{\sigma}^-}
\newcommand{\sigpop}{\hat{\sigma}^+}
\newcommand{\sigzop}{\hat{\sigma}_z}
\newcommand{\nn}{\nonumber}

\usepackage{xr-hyper}
\begin{document}

\title{Feedback-induced instabilities and dynamics in the Jaynes-Cummings model}

\date{\today}
\author{Nikolett N{\'e}met}
\email{nemet.nikolett@wigner.hu}
\affiliation{Wigner Research Centre for Physics, H-1525 Budapest, P.O. Box 49., Hungary}
\affiliation{Department of Physics,University of Auckland, Auckland, New Zealand}
\affiliation{The Dodd-Walls Centre for Photonic and Quantum Technologies, New Zealand}
\author{Scott Parkins}
\affiliation{Department of Physics,University of Auckland, Auckland, New Zealand}
\affiliation{The Dodd-Walls Centre for Photonic and Quantum Technologies, New Zealand}
\author{Victor Canela}
\affiliation{Department of Physics,University of Auckland, Auckland, New Zealand}
\affiliation{The Dodd-Walls Centre for Photonic and Quantum Technologies, New Zealand}
\author{Alexander Carmele}
\affiliation{Department of Physics,University of Auckland, Auckland, New Zealand}
\affiliation{The Dodd-Walls Centre for Photonic and Quantum Technologies, New Zealand}

%\email[Correspondence and requests for materials should be addressed to, e-mail address: ]{}

\begin{abstract}
We investigate the coherence and steady-state properties of the Jaynes-Cummings model subjected to time-delayed coherent feedback in the regime of multiple excitations. The introduced feedback qualitatively modifies the dynamical response and steady-state quantum properties of the system by enforcing a non-Markovian evolution. This leads to recovered collapses and revivals as well as non-equilibrium steady states when the two-level system (TLS) is directly driven by a laser. The latter are characterized by narrowed spectral linewidth and diverging correlation functions that are robust against the time delay and feedback phase choices. These effects are also demonstrated in experimentally accessible quantities such as the power spectrum and the second-order correlation function $g^{(2)}(\tau)$ in standard and widely available photon-detection setups.
\end{abstract}

%\setboolean{displaycopyright}{true}
\pacs{<pacs codes>}

\maketitle

%%%%%%%%%%%%%%%%%%%%%%%%%%%%%%%%%%%%%%%%%%%%%%%%%%%%%%%%%%%%%%%%%%%%%%%%%%%%%
%%%%%%%%%%%%%%%%%%%%%%%%%%%%%%%%%%%%%%%%%%%%%%%%%%%%%%%%%%%%%%%%%%%%%%%%%%%%%
%\section{Introduction}
%%%%%%%%%%%%%%%%%%%%%%%%%%%%%%%%%%%%%%%%%%%%%%%%%%%%%%%%%%%%%%%%%%%%%%%%%%%%%
%%%%%%%%%%%%%%%%%%%%%%%%%%%%%%%%%%%%%%%%%%%%%%%%%%%%%%%%%%%%%%%%%%%%%%%%%%%%%
\textit{Introduction.--- }Time-delayed feedback combines the effects of information coupling back from the environment with the non-trivial dynamics introduced by the memory of the process both in classical and non-classical (coherent) systems \cite{Lakshmanan-SenthilkumarBOOK,Bernd-HinkeBOOK,Bellen-ZennaroBOOK,Scholl2008Handbook,Scholl2016Control,Lloyd2000Coherent}. In case of a short feedback, where time delay is negligible, the evolution of the system shows reduced or enhanced system-reservoir coupling, which can be modelled within a Markovian framework \cite{Gough2009Series,Combes2017SLH,Fang2017Multiple}. For longer loops, however, the non-Markovian nature of the process becomes significant, which introduces non-trivial, time-delayed dynamics \cite{Naumann2014Steady-state,Grimsmo2014Rapid,Kopylov2015Time-delayed,Kopylov2015Dissipative,Joshi2016Quantum,Zhang2017Quantum,Loos2017Force,Li2018Concepts,Fang2018Non-markovian,Carmele2019Non-markovian}. This dynamical aspect has been used for classical control in the field of nonlinear dynamics and chaos \cite{Strogatz2000BOOK,Scholl2008Handbook,Scholl2016Control}, with special focus on Pyragas-type feedback for laser dynamics based on the Lang-Kobayashi semiclassical description \cite{Lang1980External,Sano1994Antimode,Albert2011Observing,Grimsmo2014Rapid,Kreinberg2018Quantum,Holzinger2019Quantum}. In the realm of quantum optics, these dynamical features are complemented with a direct influence on the system-reservoir coupling, resulting in suppressed decoherence. The combination of non-trivial dynamics and enhanced coherence provides a wider range of control over such intrinsic quantum features as squeezing or antibunching that are potentially detectable at the system output \cite{Lu2017Intensified,Kraft2016Time-delayed,Nemet2016Enhanced}.

\begin{figure}[b!]
\centering
\vspace{-.3cm}
\includegraphics[width=.5\linewidth]{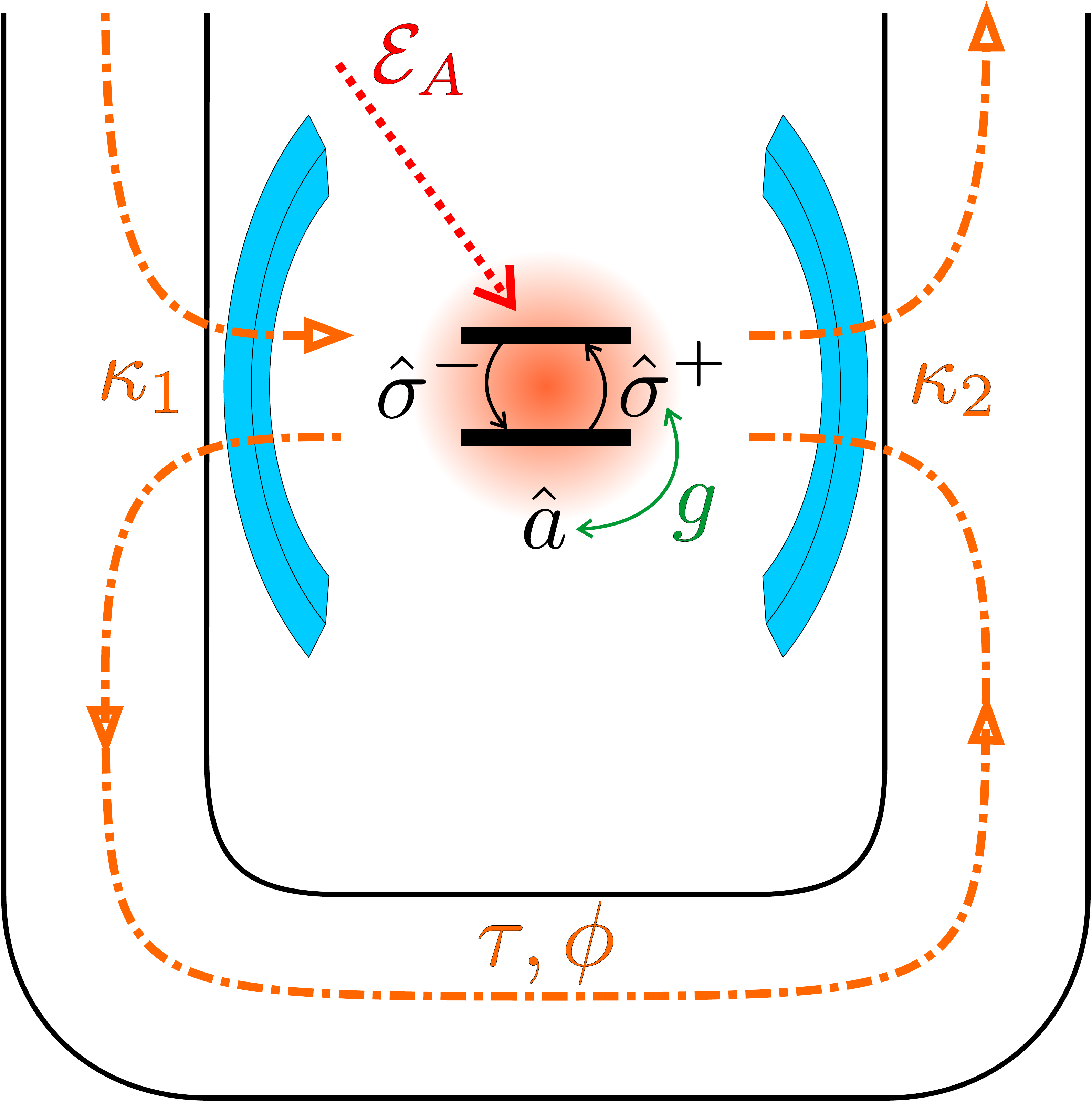}
\vspace{-.3cm}
\caption{Schematic of the setup. We consider the standard Jaynes-Cummings model, where the TLS couples to the cavity field with a strength of $g$. The waveguide field is coupled to the cavity field at two points, with respective decay rates $\kappa_{1,2}$, forming a coherent unidirectional feedback loop. A coherent driving field can also be considered for the TLS with a strength $\mathcal{E}_{\rm A}$. }
\label{fig:scheme}
\end{figure}

In the simplest case time-delayed coherent feedback (TDCF) can be realized by directly -- without any intermediate measurement -- coupling back one of the output channels of the system into one of the input channels, as shown in FIG. \ref{fig:scheme}. This structured system-reservoir coupling affects one degree of freedom in the system, and, if this is the only system variable, the dissipative dynamics leads to a fixed steady state. A classic example is the driven two-level system (TLS) in front of a mirror \cite{Dorner2002Laser-driven,Glaetzle2010Single,Pichler2016Photonic}, which has also been extensively studied experimentally \cite{Eschner2001Light,Wilson2003Vacuum,Dubin2007Photon,Andersson2019Non-exponential}. %or a chain of two-levels systems in a waveguide-QED setup \cite{Calajo2016Atom,Cascio2019Dynamics,Calajo2019exciting,Sinha2020Non-markovian,Carmele2020Pronounced}.
Probing the TLS in this setup with a coherent excitation shows feedback-induced peaks in the power spectrum as well as enhanced or reduced bunching or anti-bunching, which are sensitive to the exact value of the feedback phase. These properties are related to the entanglement building up between system, feedback loop and reservoir \cite{Pichler2016Photonic}. %The non-Markovian characteristics of this setup was also investigated in multiple experiments \cite{Eschner2001Light,Wilson2003Vacuum,Dubin2007Photon,Andersson2019Non-exponential}.%They found that the non-Markovian character of the time evolution manifests as periodic peaks in the spectrum where the spacing is inversely proportionate with the feedback delay, which is similar to the findings of \cite{Nemet2016Enhanced, Nemet2019Comparison}.

As soon as an enhanced and localized interaction is introduced between light and matter, such as in a cavity, where only the optical field is affected by feedback, signatures of more complex long-time dynamics, such as persistent oscillations, have been shown \cite{Carmele2013Single,Kabuss2015Analytical,Kraft2016Time-delayed,Nemet2016Enhanced,Nemet2019Comparison,Crowder2020Quantum}. These solutions are related to the internal coherent dynamics of the system that is protected from the intrinsically dissipative nature of TDCF and, thus, can be enhanced with the help of its coherence-recovering properties. This, however, so far has only been demonstrated in the single-excitation or linear regime, which limits the feasibility of experimental characterization and verification. To overcome these limitations, considerable efforts have been made to develop a numerical method that enables the description of a coherently probed system \cite{Pichler2016Photonic,Grimsmo2015Time-delayed,Whalen2017Open,Chalabi2018Interaction,Fang2019FDTD,Crowder2020Quantum}.

%###############################
% #################

In this Letter, making use of one of the most well-established techniques \cite{Pichler2016Photonic}, we consider the Jaynes-Cummings model with a coherent initial photonic state or with coherent driving of the TLS. We show that the oscillatory steady-state is not unique to the single-excitation subspace. Proving the truly coherent nature of TDCF, we recover the well-known collapse-revival phenomenon in the non-driven case \cite{Eberly1980Periodic,Rempe1987Observation,Chough1996Nonlinear,Carmele2019Non-markovian}. Additionally, a considerable robustness of stabilized oscillations against the choice of the feedback phase and delay time is demonstrated in the driven scenario. The long-time dynamics is accompanied by persistent oscillations in both the first- and second-order correlation functions, with diverging correlation lengths, which translates as a linewidth narrowing in the power spectrum. In the strongly driven case a collapse-revival-type phenomenon is found with extra frequencies emerging as a result of TDCF \cite{Crowder2020Quantum}.
\textit{Model.--- }We consider the Jaynes-Cummings model with a potential, direct coherent excitation of the TLS. The Hamiltonian is a combination of three contributions; the Jaynes-Cummings closed system Hamiltonian $\Hop_{\rm JC}$, the coherent driving $\Hop_{\rm dr}$, and the system-reservoir interaction $\Hop_{\rm SR}$. All interactions are considered in the rotating-wave and dipole approximations:
\begin{align}
\label{eq:H_tot}
\Hop&=\Hop_{\rm JC}+H_{\rm dr}+H_{\rm SR},\\
\label{eq:H_JC}
\Hop_{\rm JC} 
&= 
\hbar \omega_\text{C} \adop \aop 
+ \hbar \omega_\text{A} \sigzop
+ \hbar g 
\left( \adop \sigmop + \sigpop\aop \right), \\
\label{eq:H_drive}
\Hop_{\rm dr}
&=
\hbar \mathcal{E}_\text{A}
\lk\sigpop e^{-i\omega_\text{L} t} + \sigmop e^{i\omega_\text{L} t}\rk, \\
\label{eq:H_SR}
\Hop_\text{\rm SR}
&=
\hbar
\int \hspace{-.1cm}
\lka
\omega \bdop_\omega \bop_\omega
+i
\lsz
\gamma^*(\omega)\bdop_\omega \aop
-
\gamma(\omega)\adop \bop_\omega 
\rsz
\rka  {\rm d}\omega, 
\end{align}
% ###########################
where $\aop$, $\sigmop$, $\bop_\omega$ are lowering or annihilation operators, $\omega_C$, $\omega_A$, $\omega$ are the frequencies of the cavity, TLS and the reservoir excitations, respectively, $\mathcal{E}_A$ is the driving field amplitude for the TLS, and $g$ is the coupling strength between the TLS and the cavity.
In the following, we assume resonant cavity-emitter and laser-emitter interactions ($\omega_\text{C}=\omega_\text{L}=\omega_\text{A}$). The coupling between the cavity and the reservoir becomes frequency dependent due to TDCF \cite{Nemet2019Comparison}: $\gamma(\omega) = \gamma_1 \exp[-i(\omega\tau/2-\phi_1)] + \gamma_2\exp[i(\omega\tau/2+\phi_2)]$ ($\gamma_1$ ($\gamma_2$) is the coupling strength through the left (right) mirror \cite{supplemental}. For the sake of simplicity, the free emission of the TLS, which we expect to contribute as an extra linewidth broadening, is ignored.
Moving into a frame rotating at the TLS resonance frequency % and transforming out an overall phase of $\omega\tau$ as in e.g. \cite{Kabuss2015Analytical}
we obtain %Using the unitary operator
%$\Uop(t,0)=\exp
%\lsz
%i\omega_\text{A}t
%\lk\adop \aop+\sigzop\rk
%+\int  i\omega(t-\tau/2) \bdop_\omega \bop_\omega\text{d}\omega
%\rsz
%$ we move into a frame rotating by the TLS resonance frequency:
%
%\begin{align}
%\Hop =&
%\hbar g 
%\lk \adop \sigmop + \sigpop\aop \rk
%+
%\hbar \mathcal{E}_\text{A}
%\lk\sigmop + \sigpop\rk \nn\\
%&+\hbar\frac{1}{2i}
%\int 
%\lsz
%\bdop_\omega
%\lk
%\gamma_1e^{i(\omega-\omega_\text{A})t}
%+\gamma_2e^{i(\omega-\omega_\text{A})(t-\tau)}
%e^{i\omega_\text{A}\tau}
%\rk
%a \right.\nn\\
%&\left.
%\hspace{1.5cm}-\hc
%\rsz\text{d}\omega.
%\end{align}
%%
%Introducing the Fourier-transformed bath operators;
%\begin{align}
%\bdop(t)
%= \frac{1}{\sqrt{2\pi}}\int   e^{i(\omega-\omega_\text{A})t} %\bdop_\omega\text{d}\omega,
%\end{align}
%the time-dependent Hamiltonian simplifies to
\begin{align}
\Hop(t) = &
\hbar g 
\lk \adop \sigmop + \sigpop\aop \rk
+
\hbar \mathcal{E}_\text{A}
\left(\sigpop + \sigmop\right) \\
&+i\hbar
\lka
\lsz
\sqrt{2\kappa_1}\bdop(t)+
\sqrt{2\kappa_2}\bdop(t-\tau)
e^{i\phi}
\rsz
\aop\right.\nn\\
&\hspace{1cm}\left.-\hc
\rka,
\end{align}
where we use the Fourier transformed reservoir operator $\bdop(t)
= \frac{1}{\sqrt{2\pi}}\int   e^{i[(\omega-\omega_\text{A})(t+\tau/2)-\phi_1]} \bdop_\omega\text{d}\omega$, and
$\sqrt{2\kappa_j}=\gamma_j\sqrt{2\pi}$. The feedback phase, $\phi={\rm mod}_{2\pi}\lk\omega_A\tau+\phi_2-\phi_1\rk$, describes the phase relationship between the returning and emitted field at mirror 1.

This Hamiltonian considers a feedback reservoir that couples to the system at two different times. In other words, a memory is introduced for the vast environment that supports non-Markovian system dynamics. In some sense, the feedback loop together with the original system constitutes an effective coherent quantum system. Due to the fundamental issue of keeping track of the system excitations through a vast environment, numerical methods have been introduced based on various approximations to overcome this limitation \cite{Pichler2016Photonic,Grimsmo2015Time-delayed,Whalen2017Open,Chalabi2018Interaction,Fang2019FDTD,Crowder2020Quantum}. One of the most efficient techniques represents the system and reservoir states together at various times as a Matrix Product State (MPS) \cite{Schollwock2011Density}. Following Ref. \cite{Pichler2016Photonic} we model the dynamics by representing the state of the system at a given time and the state of the reservoir in short timebins together as an MPS $(\ket{\Psi(t)})$. Then, using the quantum stochastic Schr\"odinger equation approach \cite{Gardiner-ZollerBOOK}, each time step in the evolution is obtained by acting with the discretized unitary ${\rm d}\Uop(t)={\rm exp}\lk -\frac{i}{\hbar}\int_t^{t+{\rm d}t}\Hop(t^\prime){\rm d}t^{\prime}\rk$ on the wave function $\ket{\Psi(t+{\rm d}t)}={\rm d}\Uop(t)\ket{\Psi(t)}$ \cite{Pichler2016Photonic,supplemental}.
%
%Due to the strongly entangled system-reservoir dynamics,
%typical perturbative approaches fail to simulate this Hamiltonian correctly, therefore, the MPS method \cite{Pichler2016Photonic} is employed.
%

%%%%%%%%%%%%%%%%%%%%%%%%%%%%%%%%%%%%%%%%%%%%%%%%%%%%%%%%%%%%%%%%%%%%%%%%%%%%%
%%%%%%%%%%%%%%%%%%%%%%%%%%%%%%%%%%%%%%%%%%%%%%%%%%%%%%%%%%%%%%%%%%%%%%%%%%%%%
%\section{Transient dynamics without driving}
%%%%%%%%%%%%%%%%%%%%%%%%%%%%%%%%%%%%%%%%%%%%%%%%%%%%%%%%%%%%%%%%%%%%%%%%%%%%%
%%%%%%%%%%%%%%%%%%%%%%%%%%%%%%%%%%%%%%%%%%%%%%%%%%%%%%%%%%%%%%%%%%%%%%%%%%%%%
\begin{figure}[t!]
\centering
\includegraphics[width=.5\textwidth]{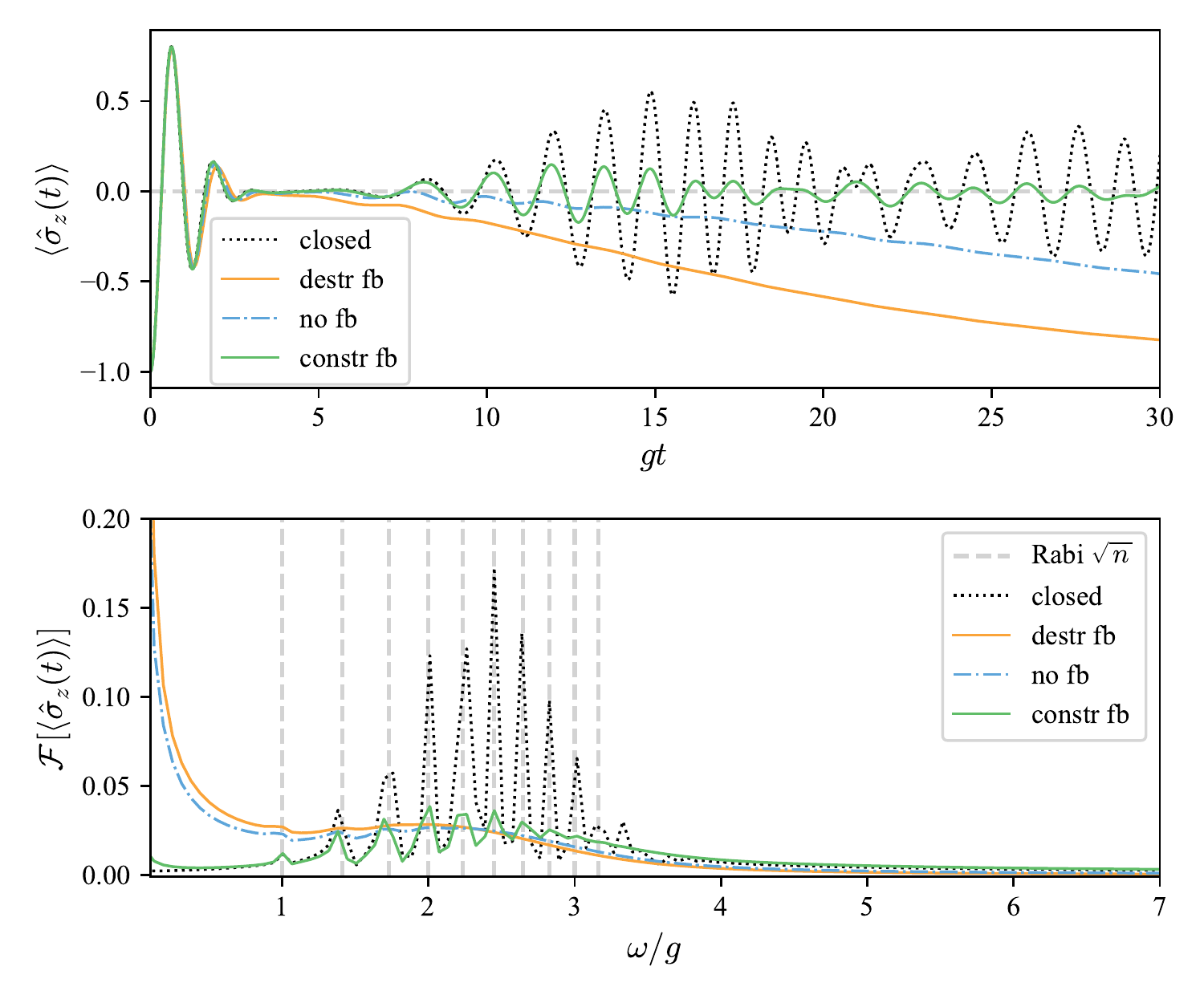}
\vspace{-1cm}
\caption{Suppressed and recovered revivals in the TLS population inversion (upper panel) and the Fourier transform (lower panel) as a result of destructive and constructive feedback, respectively. Rabi frequencies are shown with dashed grey lines. $\kappa_1=g/100$, $\kappa_2=4g/100$, $g\tau=0.04$, $\left|\alpha\right|^2=6$, constr fb: $\phi=\pi$, destr fb: $\phi=0$. }
\label{fig:collapse-revival}
\vspace{-.3cm}
\end{figure}

\textit{Transient dynamics without driving.--- }One of the most prominent features of the Jaynes-Cummings model is the collapse-revival phenomenon. In this case a closed system is considered with the cavity field initialized in a coherent state, and the TLS in its ground state. After an initial destructive interference-induced collapse, revivals of the cavity and TLS populations can be observed as the coherence in the system enables rephasing. This phenomenon stems from the uniquely quantum mechanical nature of the system and the strong coupling between cavity and TLS \cite{Shore1993Jaynes}.

%\onecolumngrid
%\twocolumngrid

Opening the system to its surroundings gives an overall exponentially decaying envelope to the cavity and TLS populations. In order to demonstrate qualitative change as a result of coherent feedback, we choose a regime where no revival can be observed without feedback (blue dash-dotted line in FIG. \ref{fig:collapse-revival}). Introducing constructive feedback in this case (destructive interference between the returning and emitted field, i.e. $\phi=\pi$) recovers a similar evolution (green solid) as expected for a closed system (black dotted line), with partial revivals. Meanwhile, a destructive feedback (constructive interference at the point of interaction $(\phi=0)$) accelerates the population damping. This finding is further emphasized by the Fourier transform of the time trace, where the distinct peaks representing the Rabi frequencies of the closed system \cite{Shore1993Jaynes} become more (less) pronounced as a result of constructive (destructive) feedback \cite{supplemental}.
% ###############################
% ###############################
% ###############################

%%%%%%%%%%%%%%%%%%%%%%%%%%%%%%%%%%%%%%%%%%%%%%%%%%%%%%%%%%%%%%%%%%%%%%%%%%%%%
%%%%%%%%%%%%%%%%%%%%%%%%%%%%%%%%%%%%%%%%%%%%%%%%%%%%%%%%%%%%%%%%%%%%%%%%%%%%%
%\section{Steady-state properties with weaker TLS driving}
%%%%%%%%%%%%%%%%%%%%%%%%%%%%%%%%%%%%%%%%%%%%%%%%%%%%%%%%%%%%%%%%%%%%%%%%%%%%%
%%%%%%%%%%%%%%%%%%%%%%%%%%%%%%%%%%%%%%%%%%%%%%%%%%%%%%%%%%%%%%%%%%%%%%%%%%%%%
%\textit{Steady-state properties with weaker TLS driving.--- }
With a lower number of excitations in the system, our simulation can also determine steady-state properties of the system. However, in order to get a non-trivial steady-state, we consider continuous, coherent driving of the TLS that is strong enough to give more than one excitation at a time in the system, but also weak enough for our simulation method to reliably determine the steady-state correlation functions and power spectrum.
\begin{figure}[t!]
\centering
\includegraphics[width=.5\textwidth]{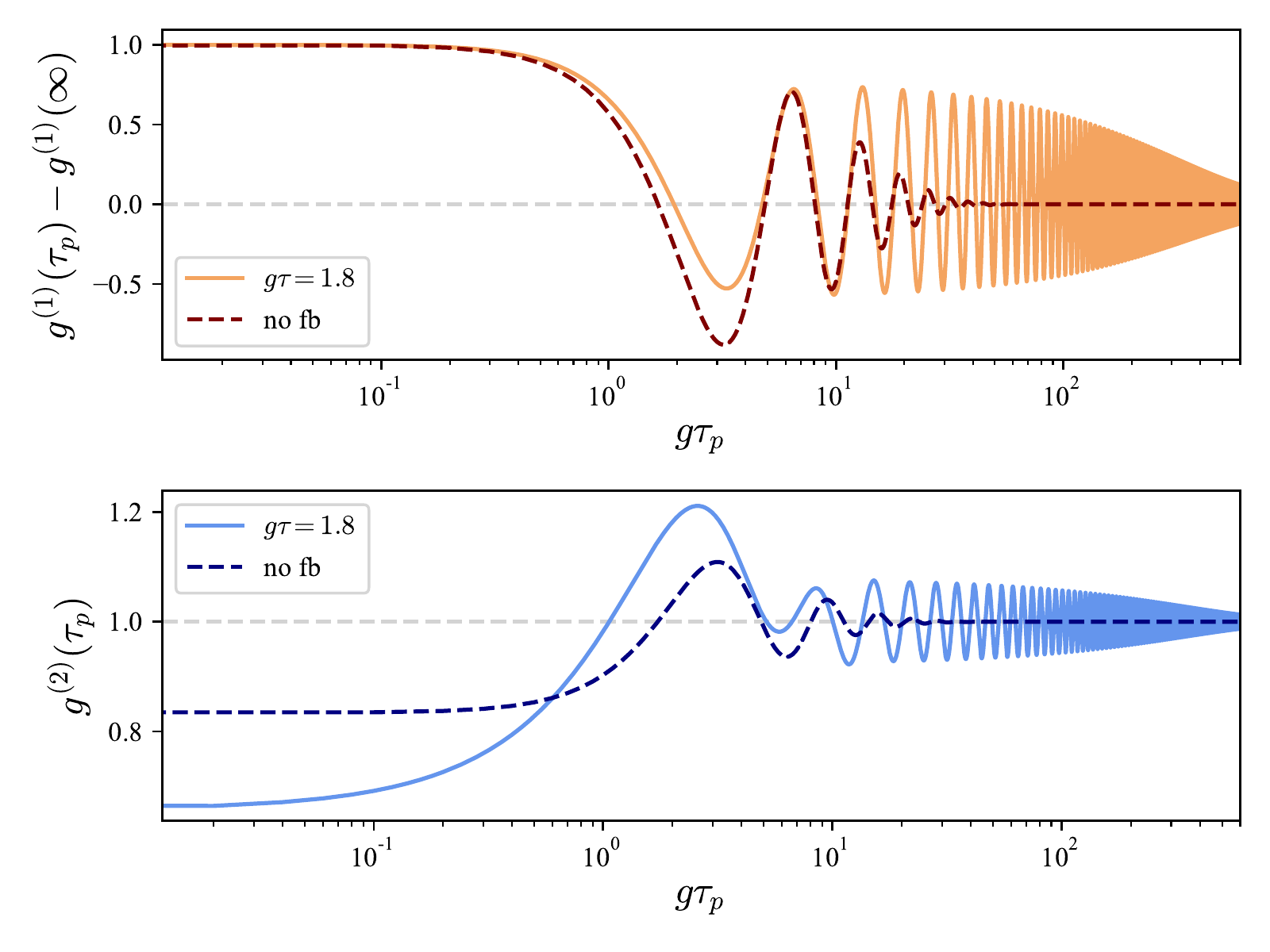}
\vspace{-1cm}
\caption{
%For certain feedback, a resonance is driven. In these cases, no steady-state is reached. The decay is only due to a finite cutoff in the SVD and finite time steps.
First-order (upper panel) and second-order (lower panel) correlation functions settling around $1$ without feedback and oscillating with feedback. $\phi=\pi/2, g=0.2, \mathcal{E}_\text{A}=0.01, \kappa_1/g=0.6125, \kappa_2/g=0.6, g\tau=1.8.$}
\label{fig:diverging_correlation_length}
\vspace{-.3cm}
\end{figure}

%%%%%%%%%%%%%%%%%%%%%%%%%%%%%%%%%%%%%%%%%%%%%%%%%%%%%%%%%%%%%%%%%%%%%%%%%%%%%
%%%%%%%%%%%%%%%%%%%%%%%%%%%%%%%%%%%%%%%%%%%%%%%%%%%%%%%%%%%%%%%%%%%%%%%%%%%%%
%\subsection{Stabilized Rabi oscillations}
%%%%%%%%%%%%%%%%%%%%%%%%%%%%%%%%%%%%%%%%%%%%%%%%%%%%%%%%%%%%%%%%%%%%%%%%%%%%%
%%%%%%%%%%%%%%%%%%%%%%%%%%%%%%%%%%%%%%%%%%%%%%%%%%%%%%%%%%%%%%%%%%%%%%%%%%%%%
\textit{Diverging correlation functions with TLS driving.--- }
Previous works have shown stabilization of Rabi oscillations as a result of TDCF in the single-excitation limit when the condition $g\tau+\phi=(2n+1)\pi$\ $(n\in\mathbb{Z})$ is satisfied \cite{Kabuss2015Analytical,Nemet2019Comparison}. In this Letter, we extend the scope of this work by considering coherent driving of the TLS, generating multiple excitations in the system. We choose to excite the TLS instead of the cavity as this scheme proves to be more efficient in activating the intrinsic non-linearity of the system \cite{Hamsen2017Two-photon}. Starting from the cavity vacuum and TLS ground states, the TLS excitation and cavity photon number show transient initial oscillations due to excitation exchange after turn-on of the driving field, before converging, in the case of no feedback, to a constant steady state after a time depending on the cavity loss and driving strength. With feedback, however, the time evolution can reach a limit cycle around $g\tau+\phi=n\pi\ (n\in\mathbb{Z})$, giving rise to persistent oscillations \cite{supplemental}. To connect the impact of TDCF to experimental accessible quantities, we consider the photon correlation functions of first- and second-order in the long-time limit:
\begin{align}
    g^{(1)}(\tau_p) &= \lim_{t\rightarrow\infty}\frac{\ew{\bdop(t)\bop(t+\tau_p)}}{\ew{\bdop(t)\bop(t)}},\\
    g^{(2)}(\tau_p) &= \lim_{t\rightarrow\infty}\frac{\ew{\bdop(t)\bdop(t+\tau_p)\bop(t+\tau_p)\bop(t)}}{\ew{\bdop(t)\bop(t)}^2}.
\end{align}

Without feedback these correlation functions tend to $1$ due to the coherent driving field, as can be seen in FIG. \ref{fig:diverging_correlation_length} (maroon and navy blue dashed lines). With feedback, persistent oscillations are evident in the correlation functions as well, signalling a highly non-classical output field. In the parameter regime of FIG. \ref{fig:diverging_correlation_length}, these oscillations only damp due to the imbalance of the cavity mirror transmissions resulting in an effective decay of the cavity field (orange and blue solid lines) \cite{supplemental}.

Note that as the second-order correlation function deviates from 1, this highly non-classical process cannot be described using a linear or semiclassical model \cite{supplemental} and is a result of a feedback-induced enhanced coherence in the system. The reported characteristic second-order correlation function can, in principle, be observed experimentally using a coincidence measurement on the output field \cite{Kimble1977Photon}.

Sweeping through a range of time delays while keeping the feedback phase fixed, we find that the non-linear character of the delayed dynamics together with the driving ensures an increased robustness of the above described unique features against the variation of the time delay \cite{supplemental}. This is in contrast with what was observed in the case of, for example, the degenerate parametric amplifier with feedback, where the parameters had to be set precisely \cite{Nemet2016Enhanced}.

%%%%%%%%%%%%%%%%%%%%%%%%%%%%%%%%%%%%%%%%%%%%%%%%%%%%%%%%%%%%%%%%%%%%%%%%%%%%%
%%%%%%%%%%%%%%%%%%%%%%%%%%%%%%%%%%%%%%%%%%%%%%%%%%%%%%%%%%%%%%%%%%%%%%%%%%%%%
%\subsection{Power spectrum}
%%%%%%%%%%%%%%%%%%%%%%%%%%%%%%%%%%%%%%%%%%%%%%%%%%%%%%%%%%%%%%%%%%%%%%%%%%%%%
%%%%%%%%%%%%%%%%%%%%%%%%%%%%%%%%%%%%%%%%%%%%%%%%%%%%%%%%%%%%%%%%%%%%%%%%%%%%%
\begin{figure}[t!]
\centering
\includegraphics[trim={4.5cm 2.3cm 3cm 2.5cm},clip, width=.45\textwidth]{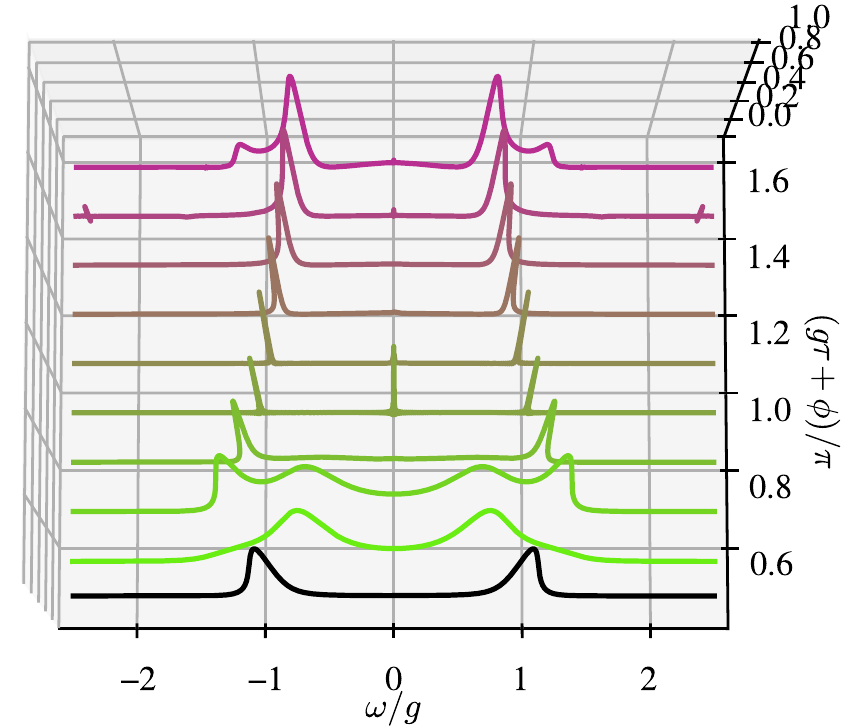}
\vspace{-.4cm}
\caption{
Power spectra for various feedback delays (coloured curves at $g\tau=0.2+0.4j, j\in\{0,...,8\}$) and without feedback (black curve). The resonances show up as Rabi splittings with narrowed linewidth and some resonant contribution. The spectra are normalized to their maximum value.
$\phi=\pi, g=0.2, \mathcal{E}_\text{A}/g=0.05, \kappa_1/g=0.6125,\kappa_2/g=0.6$}.
\label{fig:spectrum}
\vspace{-.5cm}
\end{figure}

\textit{Power spectrum.--- }
The characteristic dynamical features of the first-order correlation function can, in principle, be observed experimentally using a spectrum analyzer. The incoherent part of the obtained power spectrum is evaluated by taking the Fourier transform of $g^{(1)}$ as
\begin{align}
S(\omega) = 2\Re{\int_0^\infty \lsz g^{(1)}(\tau_p)-g^{(1)}(\infty)\rsz e^{i\omega\tau_p}    \text{d}\tau_p}.
\end{align}

Plotting power spectra over a range of time delays in FIG. \ref{fig:spectrum} \footnote{The negative values are caused by the numerics of the Fourier Transform algorithm.}, the above mentioned resonances are distinguished by narrowed linewidth at $\pm g$. Note that these sharp features can be observed over a wide range of feedback delays, which confirms the previously mentioned robustness against experimental parameter fluctuations.

The specific value of the time delay in the feedback loop has a non-trivial influence over the dominant frequencies in the dynamics. For short delays $(g\tau\approx0.6)$ the effective coupling between the cavity and the TLS is reduced, shifting the side peaks closer to resonance. As the delay increases, other peaks appear in the spectrum that can be interpreted as a result of a strong dynamical coupling between the timescale of the feedback and the cavity-TLS coupling. These spectral features are the results of the non-linear delayed dynamics and, thus, cannot be recovered by considering a linear model \cite{supplemental}.

%As observables we choose the first- and second-order photon correlation function.
%
%To compute the photon-correlation, we calculate the system- and reservoir dynamics until steady-state $t_s$.
%
% ###############################

% ###############################
%\begin{figure}
%\centering
%\includegraphics[trim={1.2cm 1.2cm 2cm 2cm},clip,width=.5\textwidth]{spectrum_w_wo_long_fb.pdf}
%\caption{System and reservoir dynamics to the steady with feedback $\tau=8$ and with feedback at $\tau=50$ . Parameters: $\phi=\pi/2, g=0.2, \mathcal{E}_\text{A}=0.01, \sqrt{\kappa}=0.35$}
%\label{fig:spectrum_long_feedback}
%\end{figure}
% ###############################

% ###############################
%\begin{figure}
%\centering
%\includegraphics[trim={0cm 0cm 0cm 0cm},clip, width=\linewidth]{unnamed.pdf}
%\caption{Power spectrum for various delay lengths. Parameters: $\phi=\pi/2, g=0.2, %\mathcal{E}_\text{A}=0.01, \sqrt{\kappa}=0.35$}
%\label{fig:spectrum_long_feedback}
%\end{figure}
% ###############################

%%%%%%%%%%%%%%%%%%%%%%%%%%%%%%%%%%%%%%%%%%%%%%%%%%%%%%%%%%%%%%%%%%%%%%%%%%%%%
%%%%%%%%%%%%%%%%%%%%%%%%%%%%%%%%%%%%%%%%%%%%%%%%%%%%%%%%%%%%%%%%%%%%%%%%%%%%%
%\subsection{Suppressed TLSic spontaneous emission}
%%%%%%%%%%%%%%%%%%%%%%%%%%%%%%%%%%%%%%%%%%%%%%%%%%%%%%%%%%%%%%%%%%%%%%%%%%%%%
%%%%%%%%%%%%%%%%%%%%%%%%%%%%%%%%%%%%%%%%%%%%%%%%%%%%%%%%%%%%%%%%%%%%%%%%%%%%%
%\textit{Suppressed TLS spontaneous emission.--- }
Focusing on the special case where the feedback phase is $\phi=\pi$, the effective dissipation rate of a symmetric cavity ($\kappa_1=\kappa_2$) approaches zero for times longer than the delay time \cite{supplemental}. In this case, the above presented condition for persistent oscillations simplifies to $g\tau=\pi$. Due to the phase difference, a destructive interference between the cavity field, the feedback, and the external driving causes suppressed excitation at the place of the TLS -- similar to that found in \cite{Alsing1992Suppression}. Therefore, in this exceptional case, the above condition also means increased mean TLS population in comparison to the cases with different delay values \cite{supplemental}. Meanwhile, the excitations in the cavity form an almost coherent field with $g^{(2)}(\infty)=1$.

% \begin{figure}[h!]
% \centering
% \includegraphics[width=.5\textwidth]{evol_pifb.pdf}
% \vspace{-1cm}
% \caption{System dynamics to the steady state with feedback at two different time delays. The destructive interference between the driving field and the cavity field leads to suppressed TLS excitation. $\phi=\pi, g=0.2, \mathcal{E}_\text{A}/g=0.05, \kappa_1/g=\kappa_2/g=0.6125$}
% \label{fig:suppressed_atom}
% \end{figure}

% ###############################
% ###############################
% ###############################
 \begin{figure}[t!]
 \centering
 \includegraphics[width=.5\textwidth]{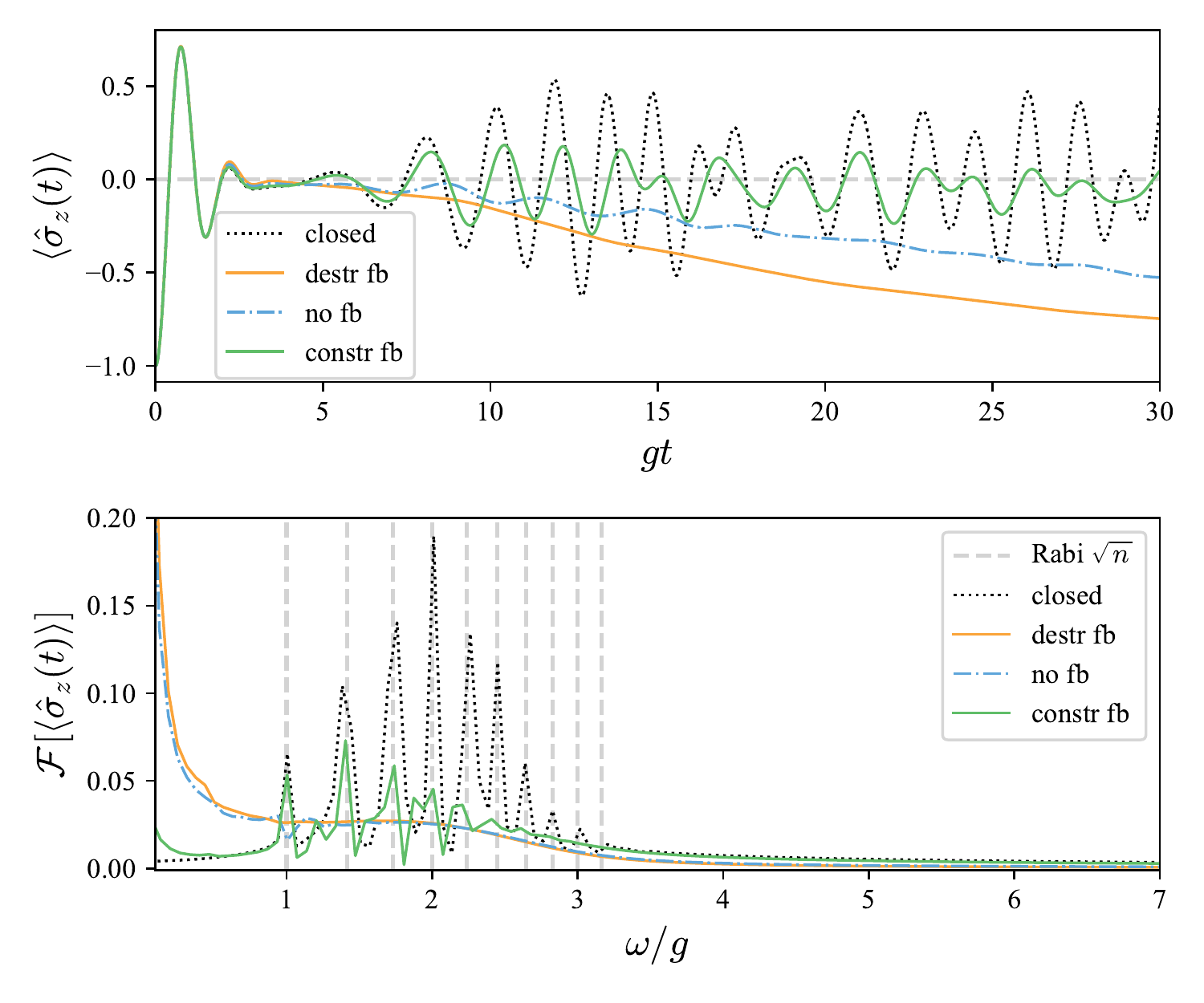}
% %\includegraphics[width=.5\textwidth]{Atomic_drive_FT.pdf}
 \vspace{-1cm}
 \caption{Time evolution of the system for strong TLS driving with and without feedback (upper panel). The Fourier transform of the dynamics with grey dashed lines representing the Rabi frequencies (lower panel). $\kappa_1=g/100$, $\kappa_2=4g/100$, $\mathcal{E}_A=2g$, $g\tau=0.04$, constr fb: $\phi=\pi$, destr fb: $\phi=0$. }
 \label{fig:strong_atom_drive}
 \vspace{-.3cm}
 \end{figure}

%%%%%%%%%%%%%%%%%%%%%%%%%%%%%%%%%%%%%%%%%%%%%%%%%%%%%%%%%%%%%%%%%%%%%%%%%%%%%
%%%%%%%%%%%%%%%%%%%%%%%%%%%%%%%%%%%%%%%%%%%%%%%%%%%%%%%%%%%%%%%%%%%%%%%%%%%%%
%\section{Transient dynamics with strong TLS driving}
%%%%%%%%%%%%%%%%%%%%%%%%%%%%%%%%%%%%%%%%%%%%%%%%%%%%%%%%%%%%%%%%%%%%%%%%%%%%%
%%%%%%%%%%%%%%%%%%%%%%%%%%%%%%%%%%%%%%%%%%%%%%%%%%%%%%%%%%%%%%%%%%%%%%%%%%%%%
\textit{Transient dynamics with strong TLS driving.--- }
Increasing the atomic driving strength, the mean photon number grows in the cavity even with no initial excitation. Introducing an imbalance in the mirror transmissions, the TLS population increases as well. Considering a regime where the system populations decay without feedback (dash-dotted blue curve in FIG. \ref{fig:strong_atom_drive}), constructive feedback ($\phi=\pi$, green solid curve) causes a similar collapse-revival as in the case of FIG. \ref{fig:collapse-revival}. Comparing the irregular revivals with the closed system dynamics at the same driving strength (black dotted curve) in FIG. \ref{fig:strong_atom_drive}, a qualitative agreement can be observed. The quasi-eigenstates of this Hamiltonian are displaced Rabi doublets \cite{Alsing1992Stark}. As such they involve a coherent cavity-field contribution supporting the emergence of revivals which become mostly dominant at large driving strengths ($\mathcal{E}_A>g$) \cite{supplemental}.

Taking the Fourier transform of these time traces, extra peaks can be observed for constructive feedback compared to the intrinsic frequencies of the closed system (lower panel of FIG. \ref{fig:strong_atom_drive}). Looking at the same with respect to the cavity field, these frequencies appear in the closed system dynamics as well \cite{supplemental}. Thus, we suggest that the extra peaks are a result of TDCF -- consisting mainly of coherent cavity field contributions -- driving the TLS. Although it is important to note that the Fourier transformation was only taken over a short time trace, signatures of such feedback-induced "half-frequencies" have also been reported for coherent cavity driving in \cite{Crowder2020Quantum}.

% ###############################
% ###############################
% ###############################

%%%%%%%%%%%%%%%%%%%%%%%%%%%%%%%%%%%%%%%%%%%%%%%%%%%%%%%%%%%%%%%%%%%%%%%%%%%%%
%%%%%%%%%%%%%%%%%%%%%%%%%%%%%%%%%%%%%%%%%%%%%%%%%%%%%%%%%%%%%%%%%%%%%%%%%%%%%
%\section{Conclusion}
%%%%%%%%%%%%%%%%%%%%%%%%%%%%%%%%%%%%%%%%%%%%%%%%%%%%%%%%%%%%%%%%%%%%%%%%%%%%%
%%%%%%%%%%%%%%%%%%%%%%%%%%%%%%%%%%%%%%%%%%%%%%%%%%%%%%%%%%%%%%%%%%%%%%%%%%%%%
\textit{Conclusion.--- }
In this Letter we investigate the effect of TDCF on the dynamical and steady-state properties of the Jaynes-Cummings model with multiple excitations. The presented characteristics are explored using an MPS-based approach in the limiting cases of high excitation and long time delay. TDCF is demonstrated to recover the well-known collapse-revival dynamics of TLS and cavity populations without driving, and causes similar TLS population dynamics in case of a strong coherent TLS driving. For weaker driving we observe persistent population oscillations that involves multiple excitations (cf. \cite{Kabuss2015Analytical,Nemet2019Comparison}) and are accompanied with oscillating, diverging first- and second-order correlation functions around $1$. The peculiar behaviour of the correlation functions strengthens the quantum mechanical origin of these features.
The presented results highlight the most crucial properties of TDCF. They show that coherence can be recovered and/or enhanced \cite{Nemet2019Stabilizing} while combining the diverse dynamical and quantum properties of a system \cite{Grimsmo2014Rapid,Nemet2016Enhanced,Kraft2016Time-delayed}. This is possible due to the the strong entanglement building up between part of the environment -- the feedback loop -- and the system. The reported striking behaviour in the observables can also be experimentally verified using common spectroscopic and coincidence measurements.

\textit{Acknowledgements. --- }We would like to thank Andr{\'a}s Vukics and P{\'e}ter Domokos for stimulating discussions. We are also grateful to the New Zealand eScience Infrasctructure (NeSI) for providing the high-performance resources for our numerical calculations. AC gratefully acknowledge support from the Deutsche Forschungsgemeinschaft (DFG) through the project B1 of the SFB 910, and from the European Union’s Horizon 2020 research and innovation program under the	SONAR grant agreement no. [734690].

\bibstyle{apsrev4-1}
\bibliography{citations}
% ###############################
% ###############################
% ###############################

\end{document}

\beginsupplement
\title{Supplemental Material to Feedback-induced instabilities and transient dynamics in the Jaynes-Cummings model}

\date{\today}
\author{Nikolett N{\'e}met}
\email{nemet.nikolett@wigner.hu}
\affiliation{The Dodd-Walls Centre for Photonic and Quantum Technologies, New Zealand}
\affiliation{Department of Physics,University of Auckland, Auckland, New Zealand}
\affiliation{Wigner Research Centre for Physics, H-1525 Budapest, P.O. Box 49., Hungary}
\author{Scott Parkins}
\affiliation{The Dodd-Walls Centre for Photonic and Quantum Technologies, New Zealand}
\affiliation{Department of Physics,University of Auckland, Auckland, New Zealand}
\author{Victor Canela}
\affiliation{The Dodd-Walls Centre for Photonic and Quantum Technologies, New Zealand}
\affiliation{Department of Physics,University of Auckland, Auckland, New Zealand}
\author{Alexander Carmele}
\affiliation{The Dodd-Walls Centre for Photonic and Quantum Technologies, New Zealand}
\affiliation{Department of Physics,University of Auckland, Auckland, New Zealand}
\maketitle
\vspace{-.5cm}

\tableofcontents
\makeatletter
\let\toc@pre\relax
\let\toc@post\relax
\makeatother

\section{Outline}
In this Supplemental we justify some of the claims in the main text that seem intuitive but non-trivial. In Section \ref{sec:int_Ham} we give a brief derivation of the Hamiltonian used in the main text. Then, in Section \ref{sec:MPS_alg} more details are given about the numerical method used in the simulations. Section \ref{sec:phase_var} clarifies the role of the feedback phase in the dynamics of the system.

One of the most important results of the Letter is the emergence of feedback-induced resonance-type dynamics that appear as persistent oscillations in the time evolution of the cavity and two-level system (TLS) populations, as well as in the output field (Section \ref{sec:time-evol}). It is also described by oscillatory correlation functions, which results in a narrowed linewidth in the power spectrum. This and other feedback-induced features of the power spectrum cannot be recovered using a linear semiclassical model as shown in Section \ref{sec:lin_spec}. The resonances appear periodically as the time delay in the feedback loop is varied. We demonstrate in Section \ref{sec:delays} that the emerging oscillatory behaviour of the population dynamics and the correlation functions show robustness against the variation of the time delay and phase. Finally, in Section \ref{sec:TLS_driving}, we aim to explain the origin of the peculiar collapse-revival-type dynamics in the TLS population present for stronger TLS driving in the main text.

% ###############################
\section{Derivation of the interaction Hamiltonian}\label{sec:int_Ham}
\begin{figure}[h!]
\centering
\includegraphics[width=.4\linewidth]{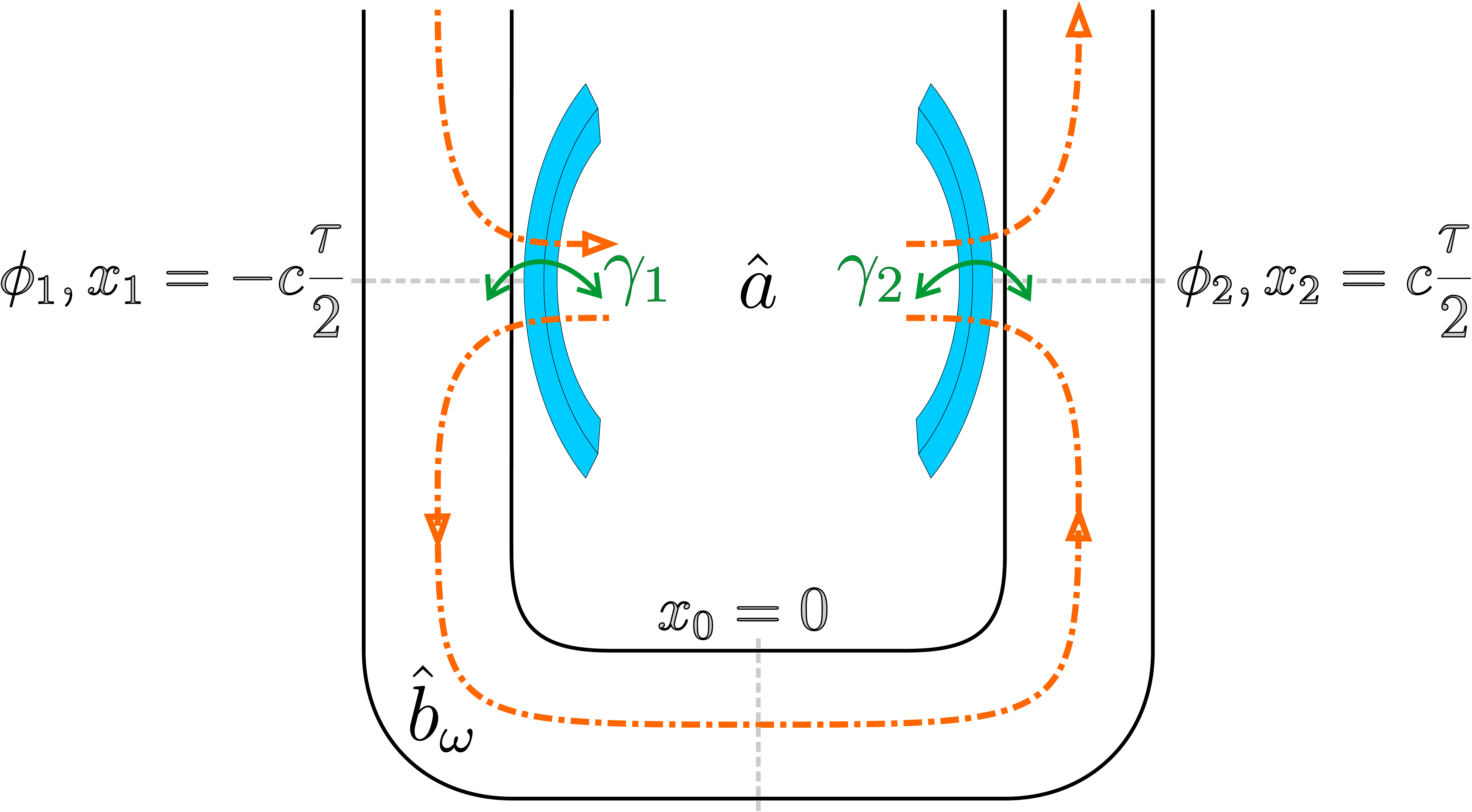}
\vspace{-.3cm}
\caption{Schematic of the cavity-waveguide coupling. Cavity mode $\aop$ couples to the propagating modes of the waveguide $\lk\bop_\omega\rk$ at two distinct points $x_1$ and $x_2$ with coupling strengths $\gamma_1$ and $\gamma_2$, respectively. }
\label{fig:coupling}
\end{figure}

% ###############################
Let us consider quantized modes of a one-dimensional waveguide with infinite length, $\bop_\omega$. The origin of our coordinate system can be chosen arbitrarily, thus, we take it to be in the middle ($x_0$ in FIG. \ref{fig:coupling}). Therefore, the points of frequency-independent coupling between the system (cavity, $\aop$) and the waveguide ($\bop_\omega$) modes are at $x_1=-c\frac{\tau}{2}$ and $x_2=c\frac{\tau}{2}$, where $\tau$ is the time delay associated with the length of the feedback loop. This implies the following coupling for plain-wave modes:
\begin{align}
\label{eq:HSR_freq}
    \Hop_{\rm SR}=&\hbar\int \lka \omega \bdop_\omega\bop_\omega-i\lk \aop+\adop\rk \lsz \gamma_1 \lk \bop_\omega e^{-i\lk\omega\tau/2-\phi_1\rk}-\bdop_\omega e^{i\lk\omega\tau/2-\phi_1\rk}\rk\right.\right.\\
     &\hspace{4.2cm}\left.\left.+    \gamma_2 \lk \bop_\omega e^{i\lk\omega\tau/2+\phi_2\rk}-\bdop_\omega e^{-i\lk\omega\tau/2+\phi_2\rk}\rk\rsz\rka {\rm d}\omega
\end{align}
where $\phi_1$ and $\phi_2$ are phase shifts obtained during the interaction (reflection or scattering).
%###############################
% #################

Using the rotating-wave approximation we obtain the system-environment interaction Hamiltonian used in the main text. This means that the overall phase shift obtained in the feedback loop is composed of the phase shifts at the point of interaction and the wave propagation.

% ###############################
\section{MPS algorithm}\label{sec:MPS_alg}
% ###############################
In the main text we defined the combined system-reservoir Hamiltonian in equations (1)-(4). Moving into an interaction picture in a frame rotating by the atomic resonance frequency, the composite wave function $\ket{\Psi(t)}$ evolves according to the Schr\"odinger equation:
\begin{align}
    \frac{{\rm d}}{{\rm d}t}\ket{\Psi(t)}=&-\frac{i}{\hbar}\Hop(t)\ket{\Psi(t)},\\
    \label{eq:HS}
\Hop(t) = &\  \hbar g \lk \adop \sigmop + \sigpop\aop \rk
+\hbar \mathcal{E}_\text{A}\left(\sigpop + \sigmop\right)&\Rightarrow&\Hop_S\\
\label{eq:HR}
&+i\hbar\lka\lsz\sqrt{2\kappa_1}\bdop(t)+\sqrt{2\kappa_2}\bdop(t-\tau)e^{i\phi}\rsz\aop-\adop\lsz\sqrt{2\kappa_1}\bop(t)+\sqrt{2\kappa_2}\bop(t-\tau)e^{-i\phi}\rsz\rka&\Rightarrow&\Hop_R.
\end{align}
In other words, as we consider both the system and reservoir in our quantum mechanical picture, the time evolution is unitary:
\begin{align}
    \ket{\Psi(t)}&=\Uop(t)\ket{\Psi(0)}\\
    \Uop(t)&=T_\leftarrow \exp\lsz-\frac{i}{\hbar}\int_0^t\Hop(t^\prime){\rm d}t^\prime\rsz
\end{align}
where $T_\leftarrow$ signals a time-ordering operation on the terms of the exponential. In order to numerically simulate the time evolution of the wave function, a coarse-graining of this unitary action is necessary as highlighted in the main text. It is important to take small enough time steps ${\rm d}t$ that the evolution is faithfully captured, but also long enough so that the correlations in the reservoir are negligible. Then
\begin{align}
    \ket{\Psi\lk t_{k+1}\rk}&={\rm d}\Uop(t_k)\ket{\Psi\lk t_k\rk}\\
    \label{eq:dUop}
    {\rm d}\Uop\lk t_k\rk&=\Uop\lk t_{k+1}\rk\Uop^\dagger\lk t_k\rk=\exp\lsz-\frac{i}{\hbar}\int_{t_k}^{t_{k+1}}\Hop(t^\prime){\rm d}t^\prime\rsz=\sum_n\frac{\lsz M_S\lk t_k\rk+M_R\lk t_k\rk\rsz^n}{n!}
\end{align}
where $t_k=k\cdot {\rm d}t, k\in\mathbb{N}$. The time integral of the reservoir operator $\bop(t)$ describes a stochastic process, more specifically a Wiener process \cite{Gardiner-ZollerBOOK,Breuer-PetruccioneBOOK}. Thus, each small integral contribution in (\ref{eq:dUop}) is associated with a Wiener increment:
\begin{align}
    \Rop(t) &= \int_0^t \bop(t^\prime){\rm d}t^\prime,\\
    \dRop(t_k) &= \Rop\lk t_{k+1}\rk -\Rop\lk t_k\rk= \int_{t_k}^{t_{k+1}}\bop(t^\prime){\rm d}t^\prime =\dRop_k.
\end{align}
As the time step is very short, we can consider a series expansion instead of the exponential such that
\begin{align}
     {\rm d}\Uop\lk t_k\rk&=\sum_n\frac{\lsz M_S\lk t_k\rk+M_R\lk t_k\rk\rsz^n}{n!}\\
     M_S\lk t_k\rk &=-\frac{i}{\hbar}\Hop_S {\rm d}t\\
     M_R\lk t_k\rk &=\lka\lsz\sqrt{2\kappa_1} \dRop_k^\dagger+\sqrt{2\kappa_2}\dRop_{k-L}^\dagger e^{i\phi}\rsz\aop-\hc\rka.
\end{align}

\begin{figure}[h!]
\centering
 \includegraphics[width=.55\linewidth]{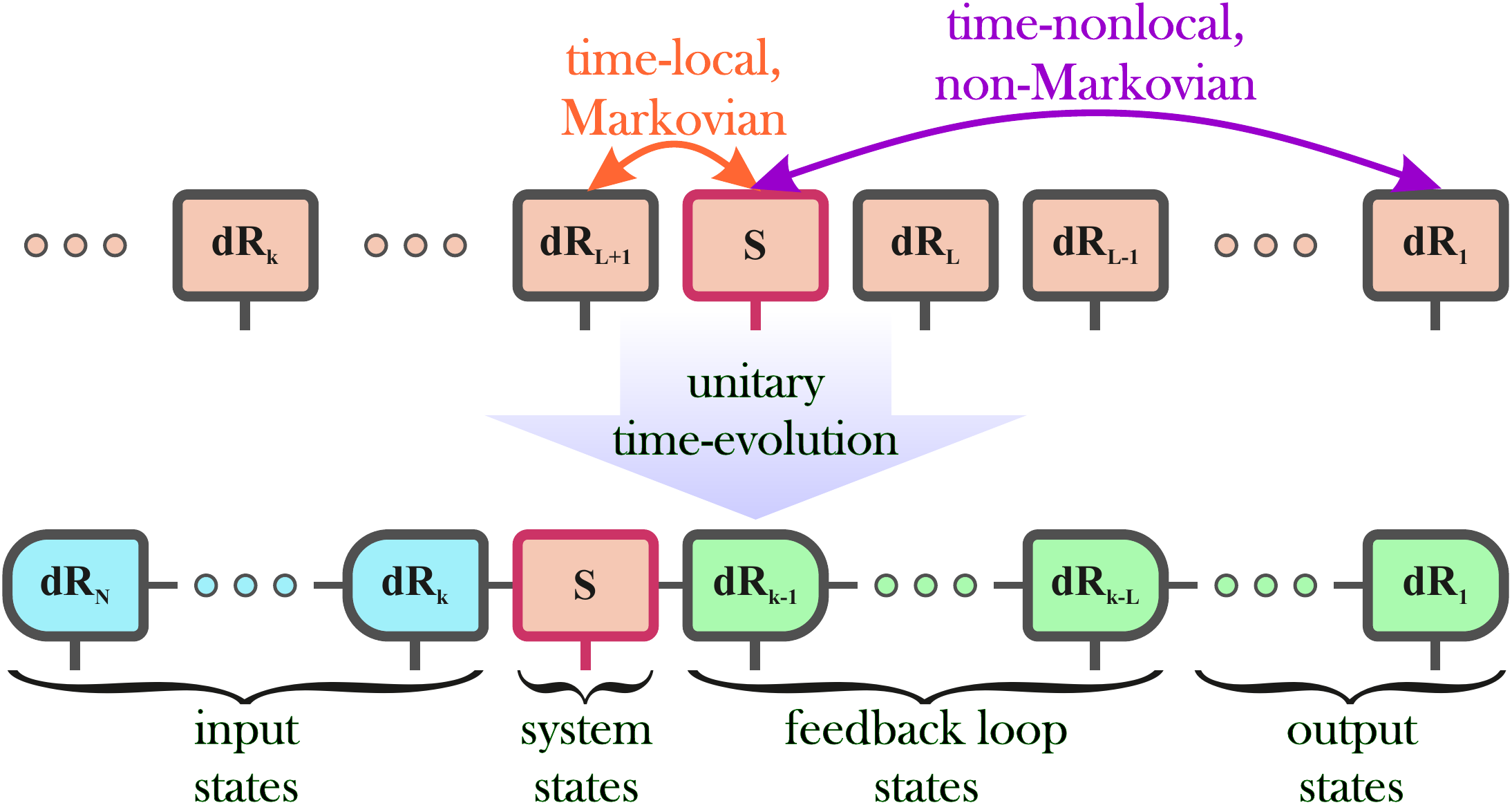}
\vspace{-.3cm}
\caption{MPS wave function before and after the time evolution. }
\label{fig:MPS}
\end{figure}

For each Wiener increment - as they are independent from each other - we define a Fock basis. Thus, the wave function is a combination of system basis functions as well as these Fock bases. An effective description of this chain can be seen in FIG. \ref{fig:MPS}. The coefficients of the many-body (system + multiple times) basis are separated by singular value decomposition into matrices. To limit the state space, only those basis functions are considered that are relevant, i.e. where the singular values are above a certain threshold.

%###############################
% #################

\begin{figure}[h!]
\centering
 \includegraphics[width=.6\linewidth]{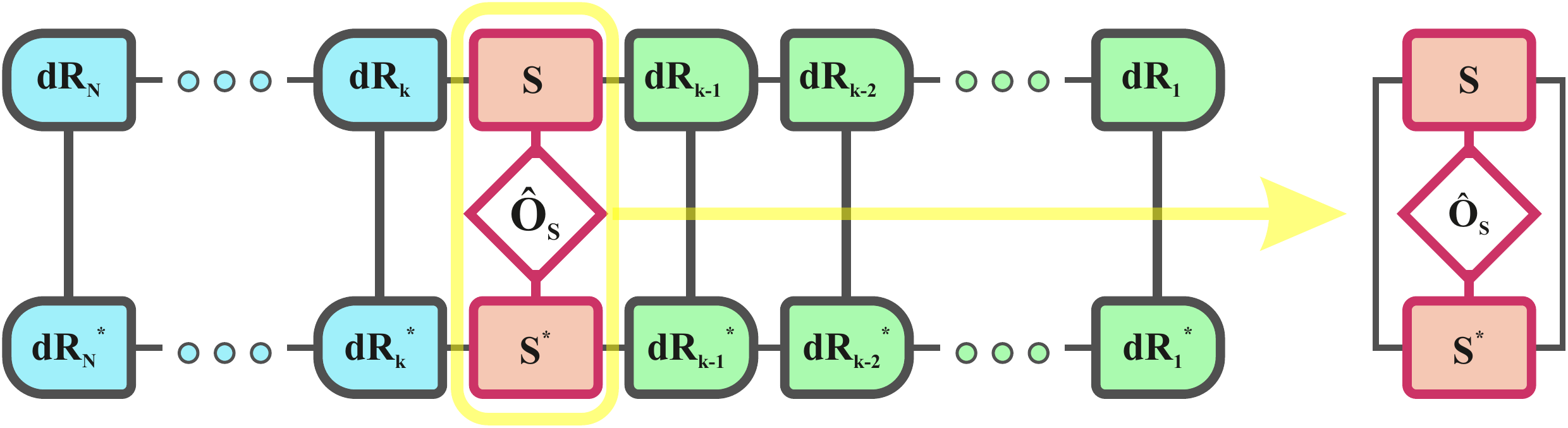}
\vspace{-.3cm}
\caption{Calculating system expectation values. }
\label{fig:sys_exp}
\end{figure}

Each bin has an extra index that corresponds to the physical (system or Fock) basis function. Thus, calculating expectation values can be done by taking the dual or conjugate of the wave function (represented by an upside-down chain in FIG. \ref{fig:sys_exp}) and contracting the indices of the observable with the corresponding physical indices.

%###############################
% #################

In each time step ($t_k$), the past bin (${\rm d}R_{k-L}$) is moved next to the system bin by SWAP operations. Afterwards the stepwise unitary acts the three bins; the present (${\rm d}R_{k}$), the system ($S$) and the past (${\rm d}R_{k-L}$) bins entangling them. Finally, the past bin is moved back. A more detailed description can be found in \cite{Pichler2016Photonic}.

\begin{figure}[h!]
\centering
 \includegraphics[width=.45\linewidth]{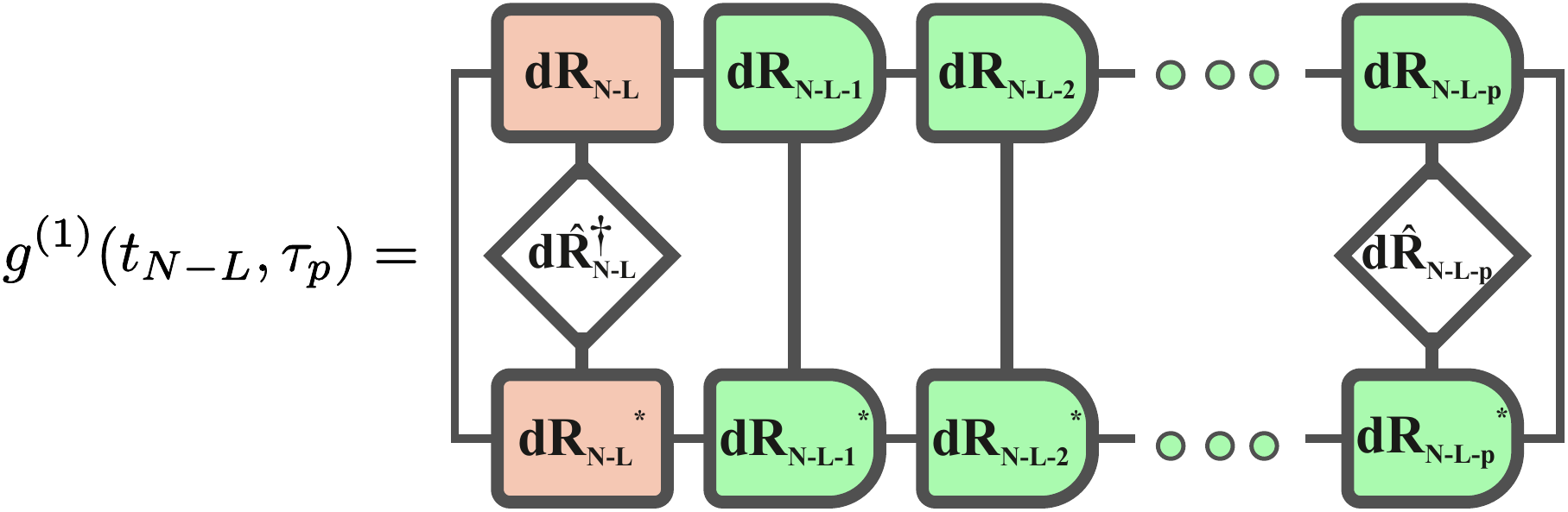}\hspace{.5cm}
 \includegraphics[width=.5\linewidth]{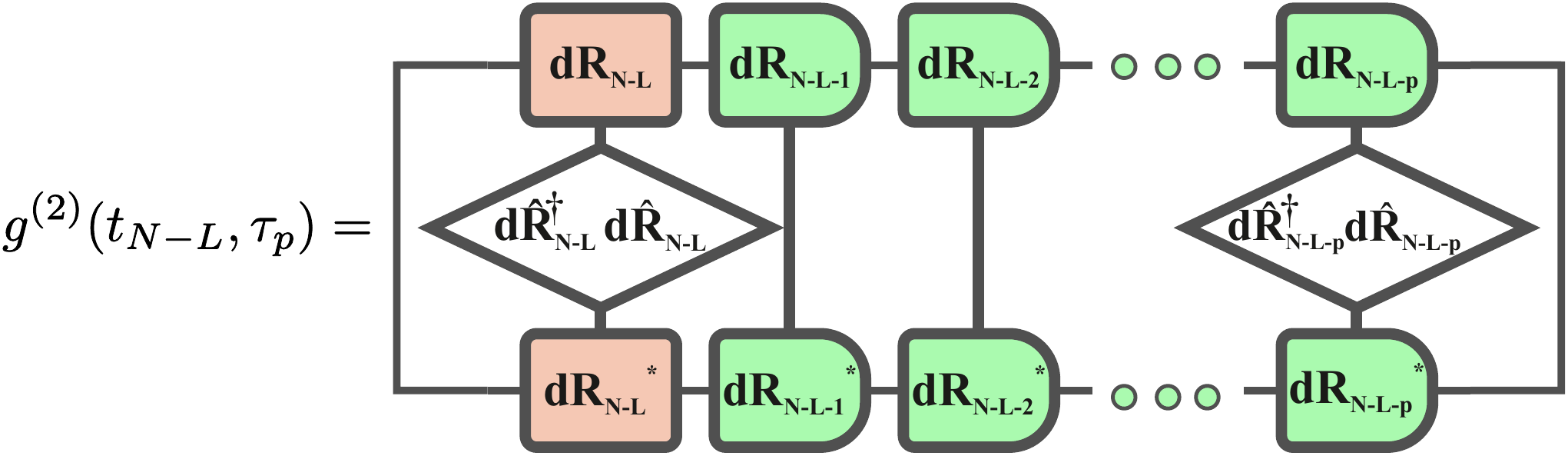}
\vspace{-.3cm}
\caption{Schematic of the cavity-waveguide coupling. Cavity mode $\aop$ couples to the propagating modes of the waveguide $\lk\bop_\omega\rk$ at two distinct points $x_1$ and $x_2$ with coupling strengths $\gamma_1$ and $\gamma_2$, respectively. }
\label{fig:correlations}
\end{figure}

The timebins with which the system already interacted with twice, can be considered the output field increments. Therefore, looking at their coherences at different times provides the first-order correlation function:
\begin{align}
    g^{(1)}(t_{N-L},\tau_p) &= \av{\dinc\Rdop(t_{N-L})\dinc\Rop(t_{N-L-p})}
\end{align}
where $N$ is the number of timebins and $\tau_p=p\dinc t, p\in\mathbb{N}$. The expectation value is evaluated similarly to the previous cases. The only difference is that the contractions with the operators are done at two different timebins (FIG. \ref{fig:correlations}). Similarly, the second-order correlation function can be evaluated as
\begin{align}
    g^{(2)}(t_{N-L},\tau_p) &= \av{\dinc\Rdop(t_{N-L})\dinc\Rop(t_{N-L})\dinc\Rdop(t_{N-L-p})\dinc\Rop(t_{N-L-p})}
\end{align}
which can be observed using coincidence measurements.
%###############################
% #################

 \noindent The other observable, the power spectrum is directly determined from the first-order correlation function:
\begin{align}
\mathcal{S}(\nu) &= \frac{2}{\dinc t}\Re\lsz\sum_{p=0}^{p_{\textrm{max}} } \lim_{t_{N-L}\to\infty}g^{(1)}(t_{N-L},\tau_p)e^{i\nu\tau_p}\rsz
\end{align}
where $p_{\rm max}$ is the maximum number of timebins over which the correlations are calculated.

% ###############################
\section{Varying the feedback phase in the Heisenberg picture}\label{sec:phase_var}
% ###############################
In order to understand the role of the phase relationship in the dynamics, it is useful to derive the Heisenberg equations of motion for the cavity field. Using the Hamiltonian (1)-(4) from the main text, we obtain the following:
\begin{align}
\label{eq:dta}
    \frac{\dinc}{\dinc t}\aop(t)&=-i\omega_C\aop(t)-ig\sigmop(t)-\int\gamma(\omega)\bop_\omega(t)\dinc\omega\\
\label{eq:dtsig}
    \frac{\dinc}{\dinc t}\sigmop(t)&=-i\omega_A\sigmop(t)+ig\sigzop\aop(t)+i\mathcal{E}_A\sigzop(t)\\
\label{eq:dtb}
\frac{\dinc}{\dinc t}\bop_\omega(t)&=-i\omega\bop_\omega(t)+\gamma^*(\omega)\aop(t)
\end{align}
Formally integrating the last one
\begin{align}
    \bop_\omega(t) &= \bop_\omega(0)e^{-i\omega t}+\gamma^*(\omega)\int_0^t\aop(t^\prime)e^{-i\omega(t-t^\prime)}\dinc t^\prime
\end{align}
and substituting it back into the equation of motion for $\aop$ we obtain:
\begin{align}
    \frac{\dinc}{\dinc t}\aop(t)&=-i\omega_C\aop(t)-ig\sigmop(t)
    -\int \gamma(\omega)\bop_\omega(0)e^{-i\omega t}\dinc\omega
    -\int|\gamma(\omega)|^2\int_0^t \aop(t^\prime)e^{-i\omega(t-t^\prime)}\dinc t^\prime\dinc\omega.
\end{align}
Let us look at the last two terms in more details:
\begin{align}
    -\int\gamma(\omega)\bop_\omega(0)e^{-i\omega t}\dinc\omega&=-\int\lsz\gamma_1e^{-i(\omega \tau/2-\phi_1)}+\gamma_2e^{i(\omega \tau/2+\phi_2)}\rsz \bop_\omega(0)e^{-i\omega t}\dinc\omega\nn\\&=\sqrt{2\kappa_1}\bop_{in}(t)+\sqrt{2\kappa_2}\bop_{in}(t-\tau)e^{i\phi^\prime}
\end{align}
where $\bop_{in}(t)=-1/\sqrt{2\pi}\int \bop_\omega(0)e^{-i[\omega(t+\tau/2)-\phi_1]}\dinc\omega$ is a $\delta$-correlated input field and $\phi^\prime=\phi_2-\phi_1$. This is where the non-Markovian nature stems from.
\begin{align}
    -\int|\gamma(\omega)|^2\int_0^t \aop(t^\prime)e^{-i\omega(t-t^\prime)}\dinc t^\prime\dinc\omega &=-\int_0^t\aop(t^\prime)\int|\gamma(\omega)|^2 e^{-i\omega(t-t^\prime)}\dinc\omega\dinc t^\prime\nn\\
    &=-\int_0^t\aop(t^\prime)\int\lsz \gamma_1^2 + \gamma_2^2 +2\gamma_1\gamma_2\cos{\lk\omega\tau+\phi^\prime\rk}\rsz  e^{-i\omega(t-t^\prime)}\dinc\omega\dinc t^\prime\nn\\
    &=-\int_0^t\aop(t^\prime)\lka 2\lk\kappa_1+\kappa_2\rk\delta(t-t^\prime)\right.\nn\\
    &\hspace{2.1cm}\left.+2\sqrt{\kappa_1\kappa_2}\lsz\delta\lk t-t^\prime+\tau\rk e^{-i\phi^\prime}+\delta\lk t-t^\prime-\tau\rk e^{i\phi^\prime} \rsz\rka\dinc t^\prime\nn\\
    &=-(\kappa_1+\kappa_2)\aop(t) - 2\sqrt{\kappa_1\kappa_2}\aop(t-\tau)e^{i\phi^\prime}
\end{align}
The first term is the normal cavity decay. The second term provides the feedback effect. If we take the zero-delay case ($\tau=0$), then the feedback phase determines the effective influence of the reservoir on the system. For a destructive feedback $\phi^\prime=2n\pi,n\in\mathbb{N}$ the effective decay rate is higher than in the without feedback case.

For constructive feedback ($\phi^\prime=(2n+1)\pi,n\in\mathbb{N}$ the effective decay rate is reduced. Note that here, we consider no loss in the feedback loop, thus for a symmetric cavity the decay is completely eliminated. The imbalance between the two sides leads to optical field leakage that is not compensated for by the destructive interference at the first mirror. Finally, if the feedback phase is $\phi^\prime=\pi/2+2n\pi, n\in\mathbb{N}$, part of the incoherent influence of the environment is converted into an effective coherent self-driving.

All these contributions can be incorporated in a Markovian description as was done in \cite{Combes2017SLH,Gough2009Series}. However, the longer the feedback loop gets, the more influence the time delay has over the dynamical features of the system. This can completely overwrite the expected decay processes as was seen e.g. in FIG. \ref{fig:suppressed_atom} where instead of suppressed atomic excitation, persistent oscillations are observed.

\section{Resonances in the time domain} \label{sec:time-evol}
\subsection{System dynamics}
\begin{figure}[h!]
\centering
\vspace{-.5cm}
\includegraphics[width=.6\textwidth]{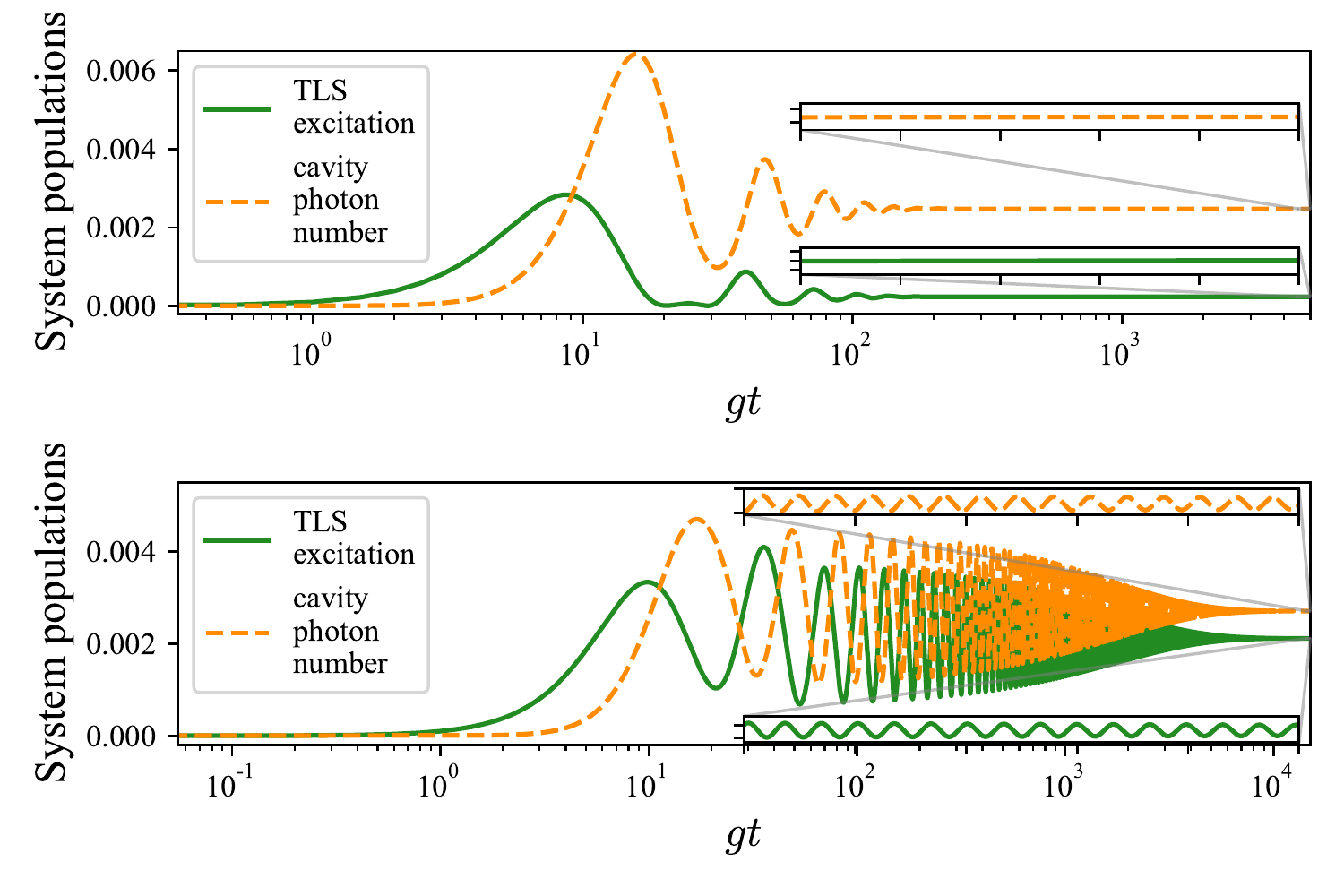}
\vspace{-.6cm}
\caption{System dynamics reaching a steady state without feedback (upper panel) and showing persistent oscillations with feedback at $g\tau=1.6$ (lower panel). $\phi=\pi/2, g=0.2, \mathcal{E}_\text{A}/g=0.05, \kappa_1/g=0.6125,\kappa_2/g=0.6$}
\label{fig:Rabi}
\end{figure}

As it is also mentioned in the main text, stabilized Rabi oscillations have been demonstrated in the single-excitation limit when $g\tau+\phi=2n\pi, n\in\mathbb{Z}$ \cite{Kabuss2015Analytical,Nemet2019Comparison}. In the present work we also see that driving the TLS with a coherent field results in a clear steady-state system population (upper panel in FIG. \ref{fig:Rabi}), which is altered by the introduction of TDCF (lower panel in FIG. \ref{fig:Rabi}).

The changing steady-state populations are a result of altered fix points due to the modified effective decay rates presented in the previous section. Varying the value of the time delay at a fixed feedback phase, however, can also introduce extra dynamical features due to emerging Hopf bifurcations \cite{Nemet2016Enhanced,Bernd-HinkeBOOK}. These are stability changes in the system and manifest as persistent oscillations in the dynamics. In subsequent sections we see that these oscillations occur periodically at certain values of the time delay and can always be described by the single Rabi frequency.

\subsection{The special case of $\phi=\pi$}
Let us present a special case of feedback phase, the so-called constructive feedback. In this case, - as was also shown in the previous section - the effective decay rate of the cavity vanishes. This resembles the scenario described in \cite{Alsing1992Suppression} where in a closed system coherent cavity field builds up that destructively interferes with the coherent TLS driving field. This results in a suppressed TLS population. 

\begin{figure}[h!]
\centering
\includegraphics[width=.6\textwidth]{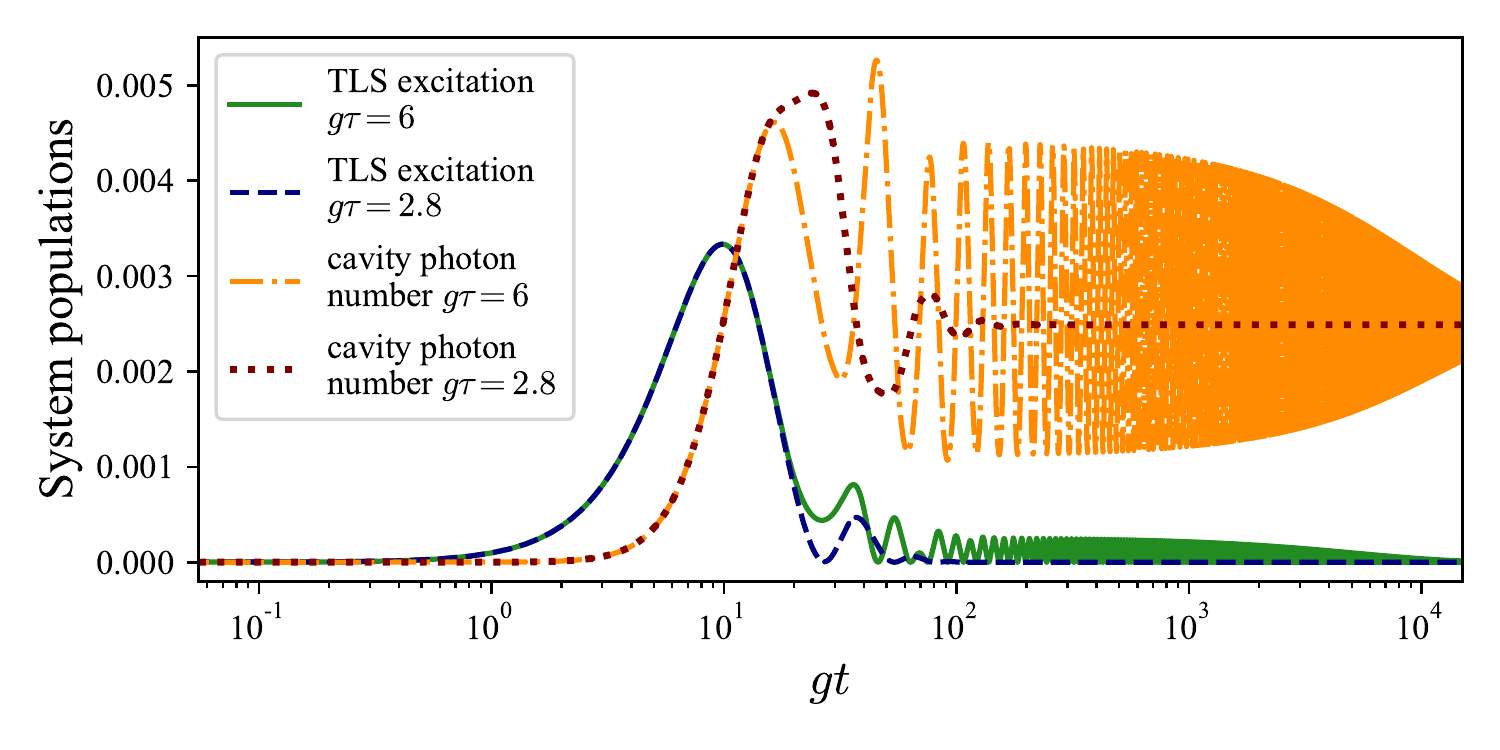}
\vspace{-.6cm}
\caption{System dynamics to the steady state with feedback at two different time delays. The destructive interference between the driving field and the cavity field leads to suppressed TLS excitation. $\phi=\pi, g=0.2, \mathcal{E}_\text{A}/g=0.05, \kappa_1/g=\kappa_2/g=0.6125$}
\label{fig:suppressed_atom}
\end{figure}

Here, the effectively closed system (after the first roundtrip in the feedback loop) gives a similarly suppressed steady-state TLS population (navy curve in FIG. \ref{fig:suppressed_atom}). Reaching a certain value of the time delay, however, can enhance this population by introducing similar persistent oscillations as before (green curve).

\subsection{Output field dynamics}\label{sec:output}
\begin{figure}[h!]
\centering
\vspace{-.5cm}
\includegraphics[width=.6\textwidth]{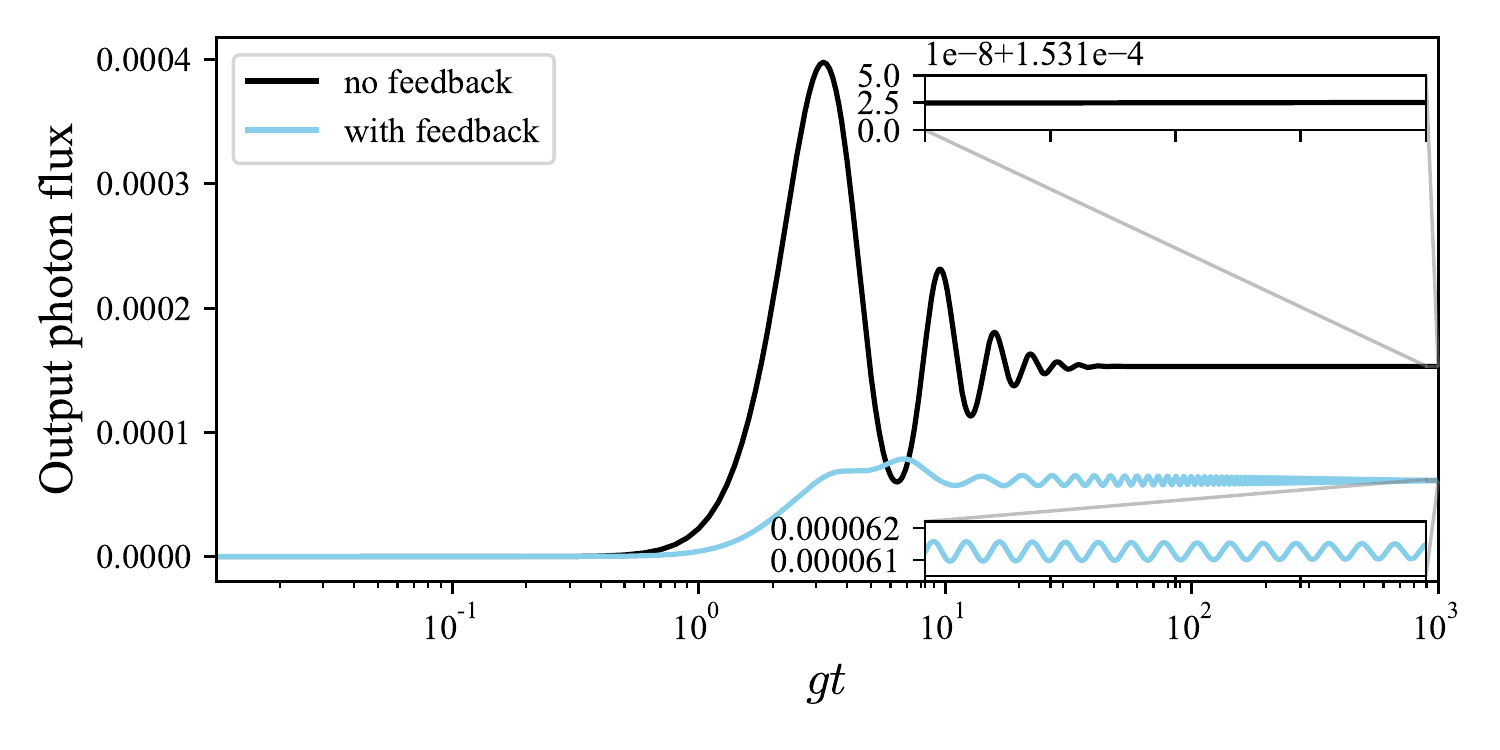}
\vspace{-.6cm}
\caption{Output photon flux corresponding to FIG. \ref{fig:Rabi}.}
\label{fig:environment}
\end{figure}
In order to show that the observed resonances are not only related to a reduced output field behaviour that recovers a hidden system dynamics, we also investigated the output photon flux of the system. In FIG. \ref{fig:environment} we see that the original steady state in the case without feedback is replaced by a similar oscillatory behaviour as in case of the cavity and atomic fields. Thus, we can conclude that the created resonance is a state that is shared between the system and the environment, a truly non-Markovian feature.

\section{Power spectra with coupled oscillators}\label{sec:lin_spec}
\begin{figure}[b!]
\centering
\includegraphics[ width=.49\textwidth]{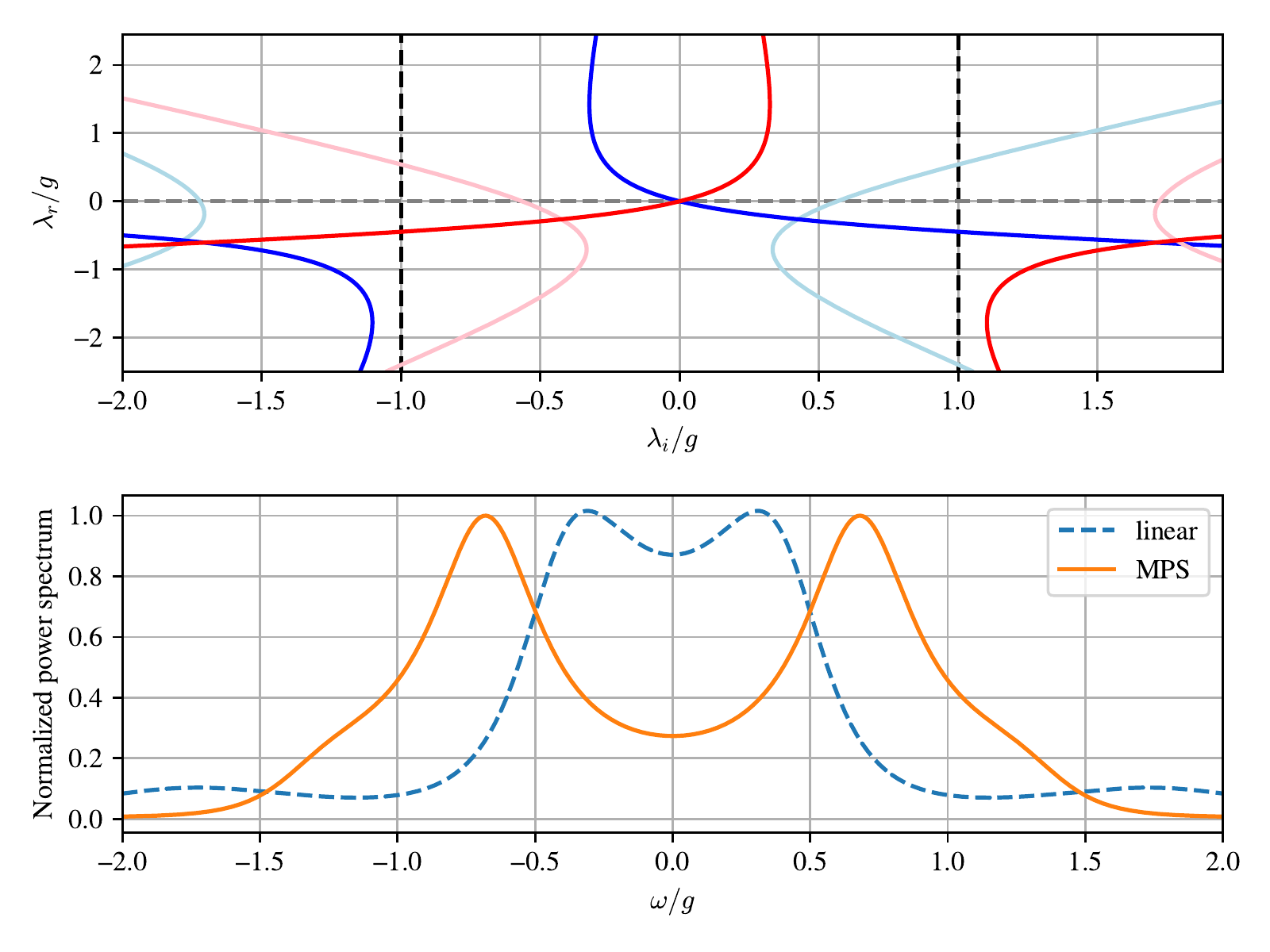}
\includegraphics[ width=.49\textwidth]{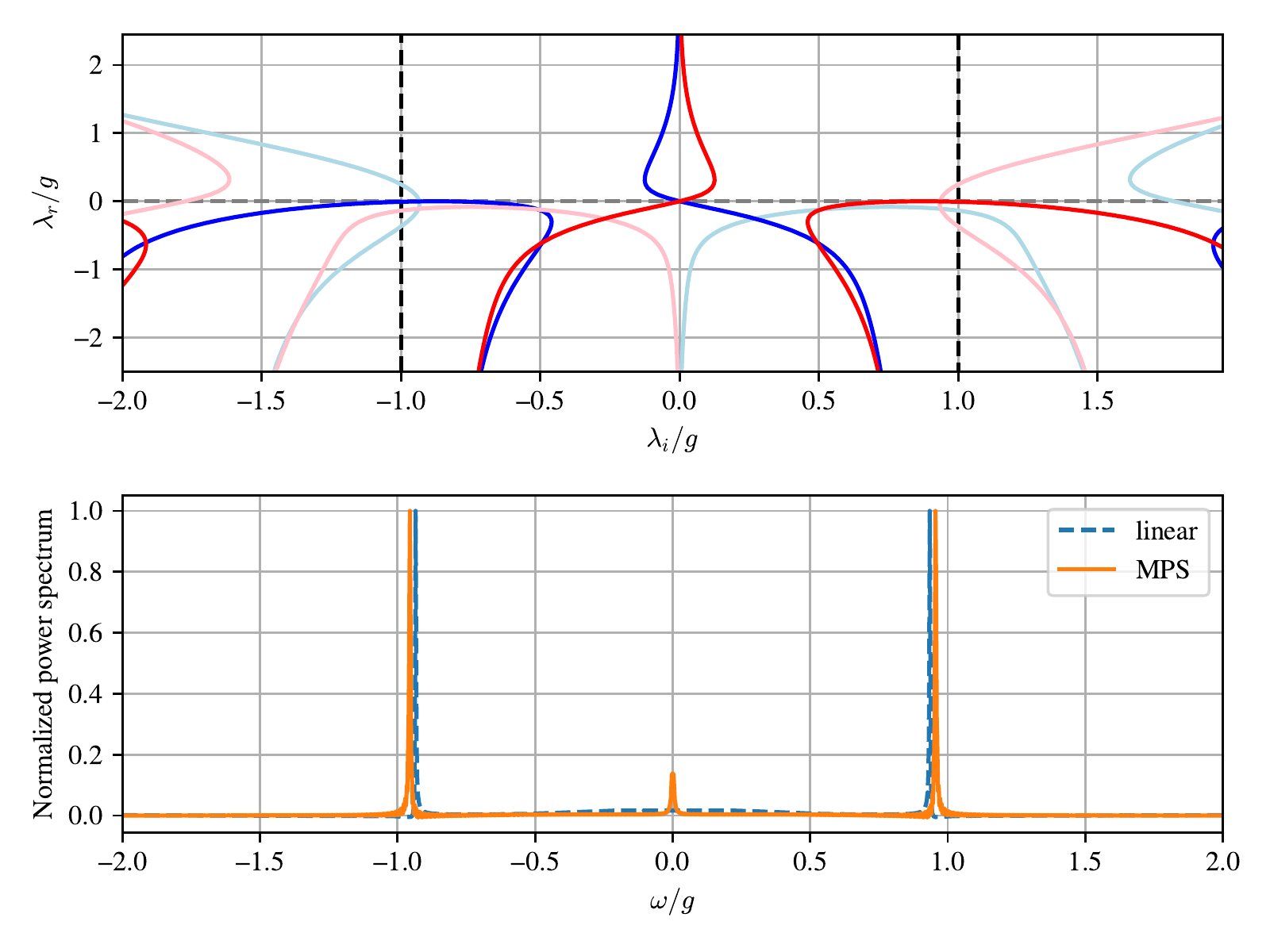}
\vspace{-.5cm}
\caption{
Graphical solution for the stability equations. Poles are located at the intersection of the red and light red as well as blue and light blue curves. The power spectrum obtained using these solutions (dashed) is compared with the peaks in the power spectrum (solid) for $g\tau=0.2$ (left) and $g\tau=1.8$ (right). Black dashed lines signal the Rabi frequencies.
Parameters: $\phi=\pi/2, g=0.2, \mathcal{E}_\text{A}/g=0.05, \kappa_1/g=0.6125,\kappa_2/g=0.6$}
\label{fig:stab-spec}
\end{figure}

Delayed equations of motion can be handled easily in the linear regime by moving into the frequency domain. In this case we assume that the atom is barely excited or that there is only a single excitation in the whole system+reservoir complex \cite{Nemet2019Comparison}. Using this approximation together with the previously determined effective role of the non-Markovian environment on the dynamics we obtain the following set of equations of motion from (\ref{eq:dta})-(\ref{eq:dtb}) in the frame rotating by the atomic resonance frequency:
\begin{align}
    \frac{\dinc}{\dinc t}\aop(t) &= -ig\sigmop(t)-\lk \kappa_1+\kappa_2+i\Delta\rk \aop(t) -2\sqrt{\kappa_1\kappa_2}\aop(t-\tau)e^{i\phi}+\sqrt{2\kappa_1}\bop_{in}(t) +\sqrt{2\kappa_2}\bop_{in}(t-\tau)e^{i\phi}\\
    \frac{\dinc}{\dinc t}\sigmop(t) &= -i\Delta\sigmop(t) -ig\aop(t)-i\mathcal{E}_A\\
    \frac{\dinc}{\dinc t}\adop(t) &= ig\sigpop(t)-\lk \kappa_1+\kappa_2-i\Delta\rk \adop(t) -2\sqrt{\kappa_1\kappa_2}\adop(t-\tau)e^{i\phi}+\sqrt{2\kappa_1}\bdop_{in}(t) +\sqrt{2\kappa_2}\bdop_{in}(t-\tau)e^{i\phi}\\
    \frac{\dinc}{\dinc t}\sigpop(t) &= i\Delta\sigpop(t) +ig\adop(t)+i\mathcal{E}_A    
\end{align}

The above single-excitation or weak-driving limit means that the TLS is also treated as a harmonic oscillator. In order to obtain the peaks in the power spectrum we perform a Laplace transform 
\begin{align}
    \asop(s) &= \int_0^\infty \aop(t)e^{-st}\dinc t\\
    \int_0^\infty \lsz\frac{\dinc}{\dinc t}\aop(t)\rsz e^{-st}\dinc t&=\lsz\aop(t)e^{-st}\rsz_0^\infty-\int_0^\infty \aop(t)(-s) e^{-st}\dinc t=s\asop(s)-\aop(0)\\
    \int_0^\infty \lsz\aop(t-\tau)\rsz e^{-st}\dinc t&=\int_{-\tau}^\infty \lsz\aop(t^\prime)\rsz e^{-s(t^\prime+\tau)}\dinc t=\asop(s)e^{-s\tau}\\
    \int_0^\infty \mathcal{E}_A e^{-st}\dinc t&=-\frac{\mathcal{E}_A}{s}\lsz e^{-st}\rsz_0^\infty=\frac{\mathcal{E}_A}{s}
\end{align}
on the equations of motion which gives:
\begin{align}
s\asop(s) &= -ig\sigmsop(s)-(\kappa_1+\kappa_2+i\Delta)\asop(s)-2\sqrt{\kappa_1\kappa_2}\asop(s)e^{-s\tau+i\phi}+\sqrt{2\kappa_1}\bsop_{in}(s)+\sqrt{2\kappa_2}\bsop_{in}(s)e^{-s\tau+i\phi}\\
s\sigmsop(s) &= -i\Delta\sigmsop(s) -ig\asop(s)-i\frac{\mathcal{E}_A}{s}\\
s\asdop(s) &= ig\sigpsop(s)-(\kappa_1+\kappa_2-i\Delta)\asdop(s)-2\sqrt{\kappa_1\kappa_2}\asdop(s)e^{-s\tau-i\phi}+\sqrt{2\kappa_1}\bsdop_{in}(s)+\sqrt{2\kappa_2}\bsdop_{in}(s)e^{-s\tau-i\phi}\\
s\sigpsop(s) &= i\Delta\sigpsop(s) +ig\asdop(s)+i\frac{\mathcal{E}_A}{s}
\end{align}
where we considered $\aop(0)=\sigmop(0)=0$. Solving these for the cavity operators gives
\begin{align}
    \asop(s)&=\underbrace{\frac{1}{s+\kappa+i\Delta+ke^{-s\tau+i\phi}-\frac{g^2}{s+i\Delta}}}_{\text{poles}}
    \lsz \underbrace{\frac{g\mathcal{E}_A}{s(s+i\Delta)}}_{\text{coherent contribution}}+\underbrace{\lk\sqrt{2\kappa_1}+\sqrt{2\kappa_2}e^{-s\tau+i\phi}\rk\bsop_{in}(s)}_{\text{incoherent contribution}}\rsz\\
    \asdop(s)&=\underbrace{\frac{1}{s+\kappa-i\Delta+ke^{-s\tau-i\phi}-\frac{g^2}{s-i\Delta}}}_{\text{poles}}
    \lsz \underbrace{\frac{g\mathcal{E}_A}{s(s-i\Delta)}}_{\text{coherent contribution}}+\underbrace{\lk\sqrt{2\kappa_1}+\sqrt{2\kappa_2}e^{-s\tau-i\phi}\rk\bsdop_{in}(s)}_{\text{incoherent contribution}}\rsz
\end{align}
The above presented denominators become poles when they approach 0. This condition provides two complex equations the real and imaginary parts of which are shown as dark and light red/blue in FIG. \ref{fig:stab-spec}. The poles are where the dark and light curves of the same colour intersect each other.

\begin{figure}[b!]
\centering
\vspace{-.5cm}
\includegraphics[trim={4.5cm 2.3cm 3cm 2cm},clip, width=.49\textwidth]{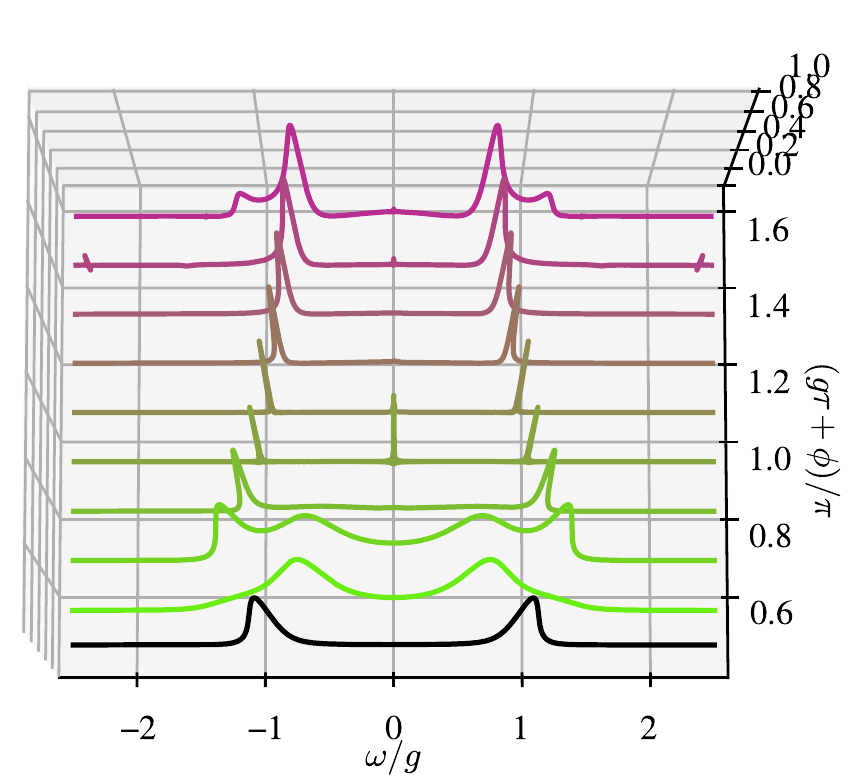}
\includegraphics[trim={4.5cm 2.3cm 3cm 2cm},clip, width=.49\textwidth]{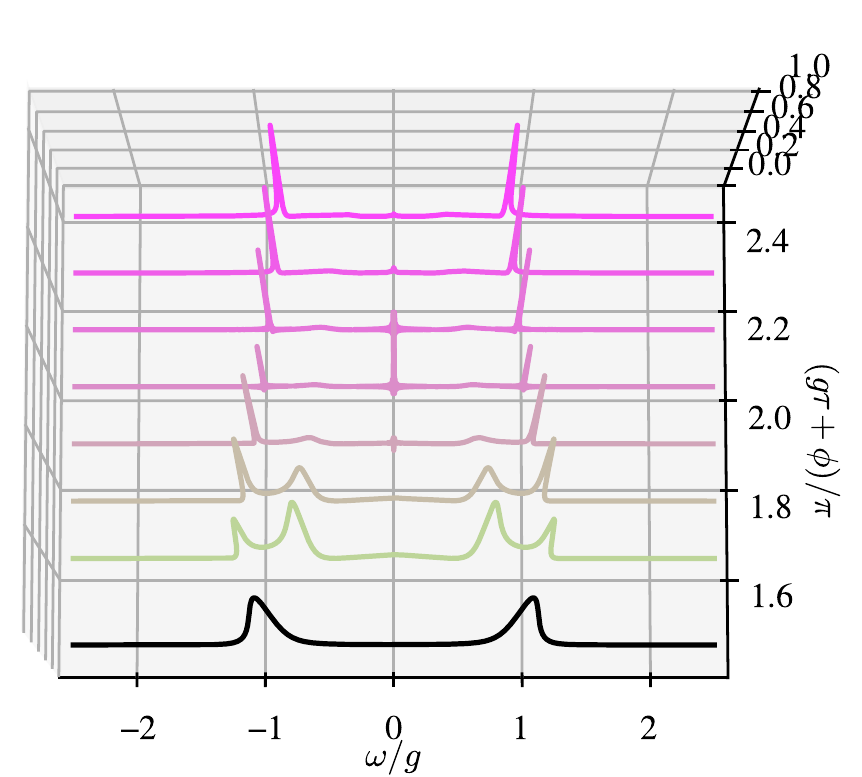}
\vspace{-.5cm}
\caption{Power spectra for various time delays}
\label{fig:tau_spectra}
\end{figure}
We can also calculate the power spectrum itself that is shown together with the MPS spectra in the bottom row of FIG. \ref{fig:stab-spec} as follows:
\begin{align}
    S(\omega)&=\av{\asdop(-i\omega)\asop(-i\omega)}-\av{\asdop(-i\omega)}\av{\asop(-i\omega)}\\
    &=\frac{g^2\mathcal{E_A}^2}{\lsz \kappa-i(\omega-\Delta)+ke^{i\lk\omega\tau+\phi\rk}-\frac{ig^2}{\omega+\Delta}\rsz\lsz \kappa-i(\omega+\Delta)+ke^{i\lk\omega\tau-\phi\rk}-\frac{ig^2}{\omega-\Delta}\rsz s^2 \lk s^2+\Delta^2\rk}
\end{align}
where $\av{\dots}$ signals vacuum expectation values. Note that the positions of the peaks of the linear model is well-predicted by the poles in the top row. The MPS spectra far from the resonance-type features, however, cannot be reconstructed using the linear approach which strengthens the point that these features are observed in a multiple-excitation regime.

For most delay values the MPS and linearised spectra give very different features. However, good agreement can be obtained close to the aforementioned resonances. 
% ###############################
\section{The effect of varying time delay}\label{sec:delays}
\begin{figure}[h!]
\centering
\includegraphics[trim={4.5cm 2.3cm 3cm 2cm},clip,width=.49\textwidth]{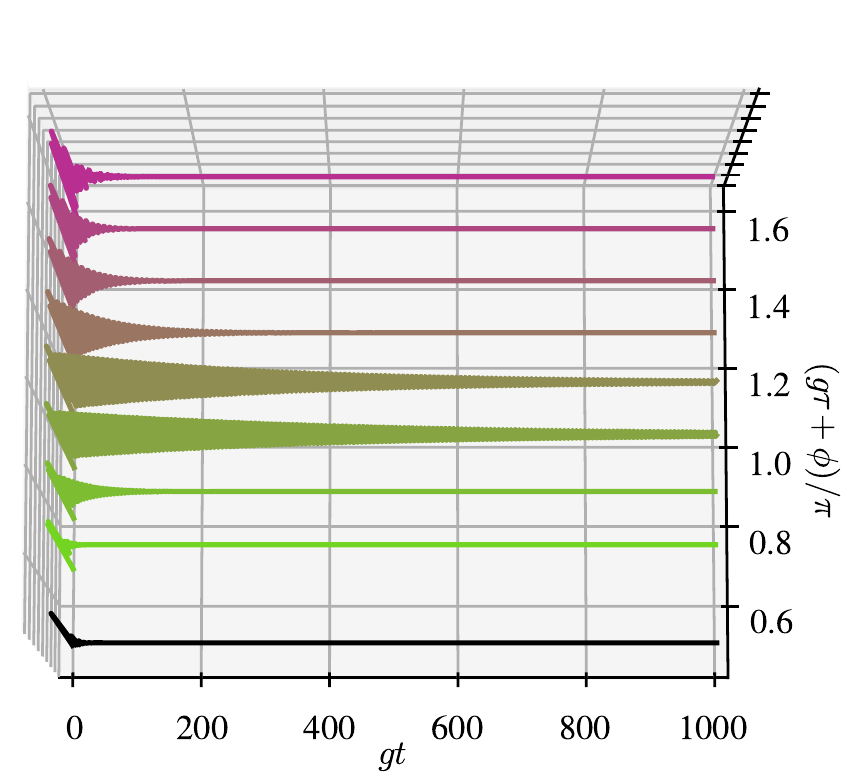}
\includegraphics[trim={4.5cm 2.3cm 3cm 2cm},clip,width=.49\textwidth]{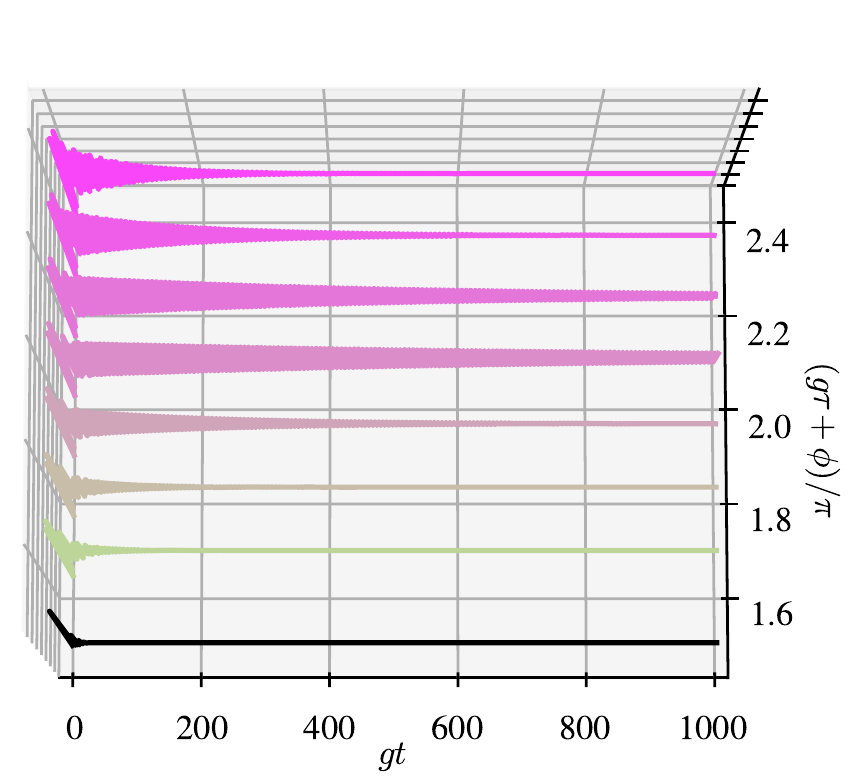}
\vspace{-.5cm}
\caption{Atomic dynamics corresponding to the spectra above.}
\label{fig:tau_dyn}
\includegraphics[trim={4.5cm 2.3cm 3cm 2cm},clip,width=.49\textwidth]{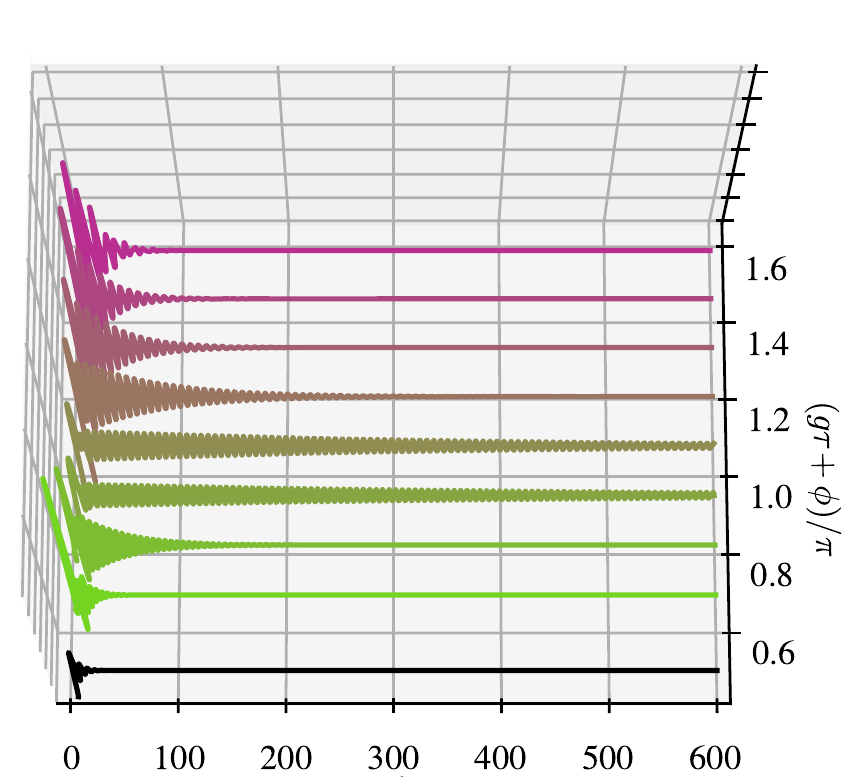}
\includegraphics[trim={4.5cm 2.3cm 3cm 2cm},clip,width=.49\textwidth]{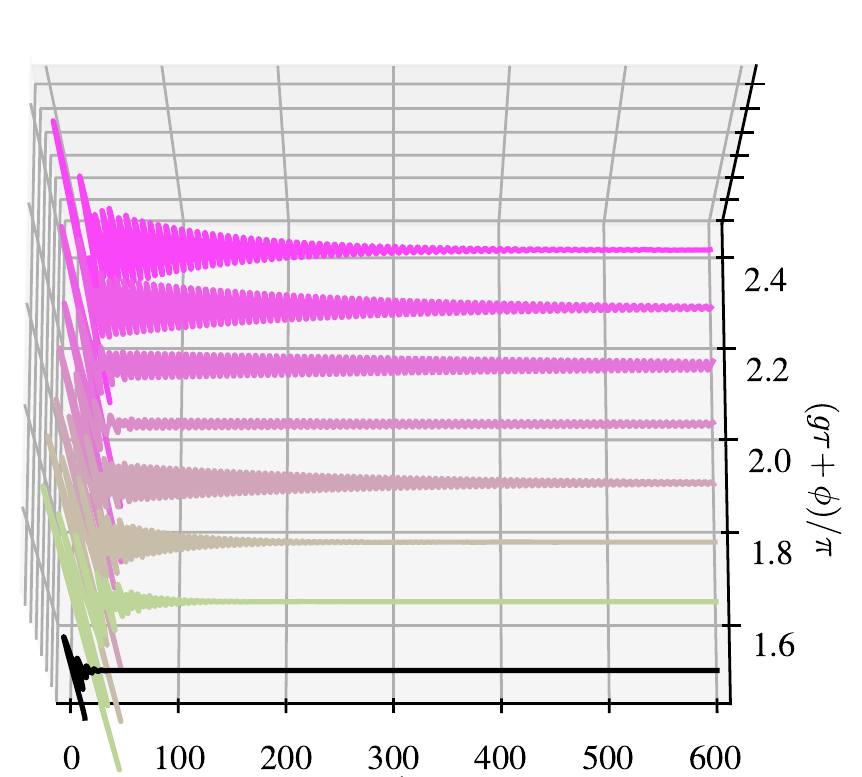}
\vspace{-.5cm}
\caption{Second order correlation functions corresponding to the spectra above.}
    \label{fig:tau_g2}
\end{figure}

% ###############################
In this section we present the same quantities as in the main part but over a larger set of time delays. FIG. \ref{fig:tau_spectra}-\ref{fig:tau_g2} summarizes these results with the help of colour matching the power spectrum, time evolution and the second-order correlation function that corresponds to the same time delay. A characteristic property of delayed dynamics is that due to the infinite number of stability eigenvalues some features are periodically reoccurring when varying the time delay. A good example of such a phenomenon is the resonance-type behaviour mentioned in the main text.

There are two time delays in this set where such dynamics occur at $g\tau+\phi=\pi$ and $2\pi$. These points are described by three very well-defined sharp peaks in the power spectra (FIG. \ref{fig:tau_spectra}) that originate from the long-living oscillations in the time domain shown in FIG. \ref{fig:tau_dyn}, respectively. Interestingly, the corresponding second-order correlation function also shows oscillations around the steady-state value of $1$ (FIG. \ref{fig:tau_g2}).

% \subsection{Weak driving}

%\begin{figure}[h!]
%\centering
%\includegraphics[trim={1.2cm 1.2cm 2cm 2cm},clip,width=.5\textwidth]{spectrum_w_wo_fb.pdf}
%\includegraphics[trim={1.2cm 1.2cm 2cm 2cm},clip,width=.5\textwidth]{g2_w_wo_fb.pdf}
%\caption{Second-order correlation function without feedback and with feedback at $g\tau=3$. Parameters: $\phi=\pi/2, g=0.2, \mathcal{E}_\text{A}=0.01, \sqrt{\kappa}=0.35$}
%\label{fig:spectrum_g2_w_wo_feedback}
%\end{figure}

\clearpage
% ###############################
\section{Characteristic dynamics with increasing TLS driving}\label{sec:TLS_driving}

\subsection{Second-order correlation function}
% ###############################
Besides the dynamics of the system populations in FIG. 5, we also looked at the instantaneous second-order correlation function $g^{(2)}(0,t)$ over the time evolution.
Initially the output field is antibunched as the atom can only accept a single excitation from the coherent driving field. After reaching the timescale of the cavity-atom dynamics, the output signal settles around a coherent response, as the cavity acts as a buffer for these atomic excitations.

Around $gt=10$, a slight bunching seems to appear that settles down to an almost coherent signal for the destructive and no-feedback cases. This can be a signature of a photon blockade breakdown-type evolution \cite{Carmichael2015Breakdown} that generates stronger bunching in the constructive feedback case where the photons returning to the cavity interfere constructively with the excitations leaving.

\begin{figure}[h!]
\centering
\includegraphics[width=.6\textwidth]{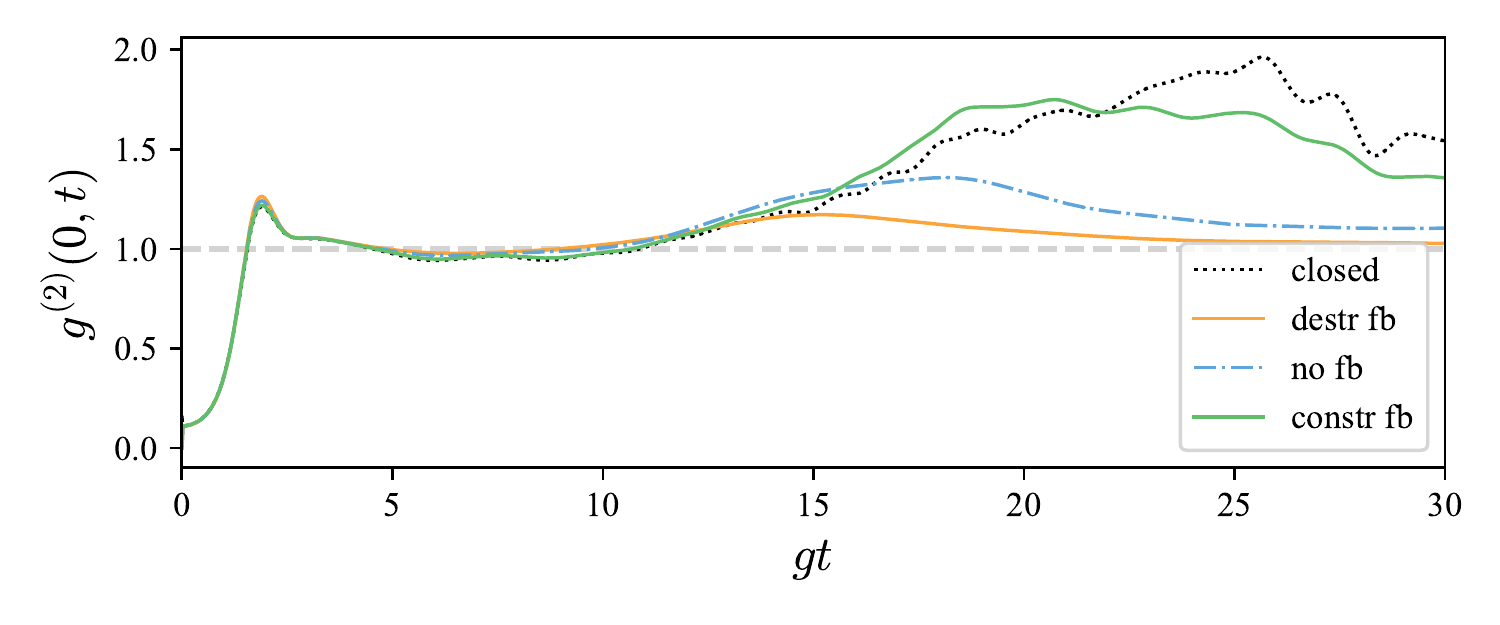}
\vspace{-.7cm}
\caption{Second-order correlation function without feedback and with feedback at $g\tau=3$. Parameters: $\kappa_1=g/100$, $\kappa_2=4g/100$, $g\tau=0.04$, $\epsilon_0=0.5g$, constr fb: $\phi=\pi$, destr fb: $\phi=0$.}
\label{fig:spectrum_g2_w_wo_feedback}
\end{figure}

\subsection{Time evolution of the closed system}
\begin{figure}[h!]
\centering
\includegraphics[width=.7\textwidth]{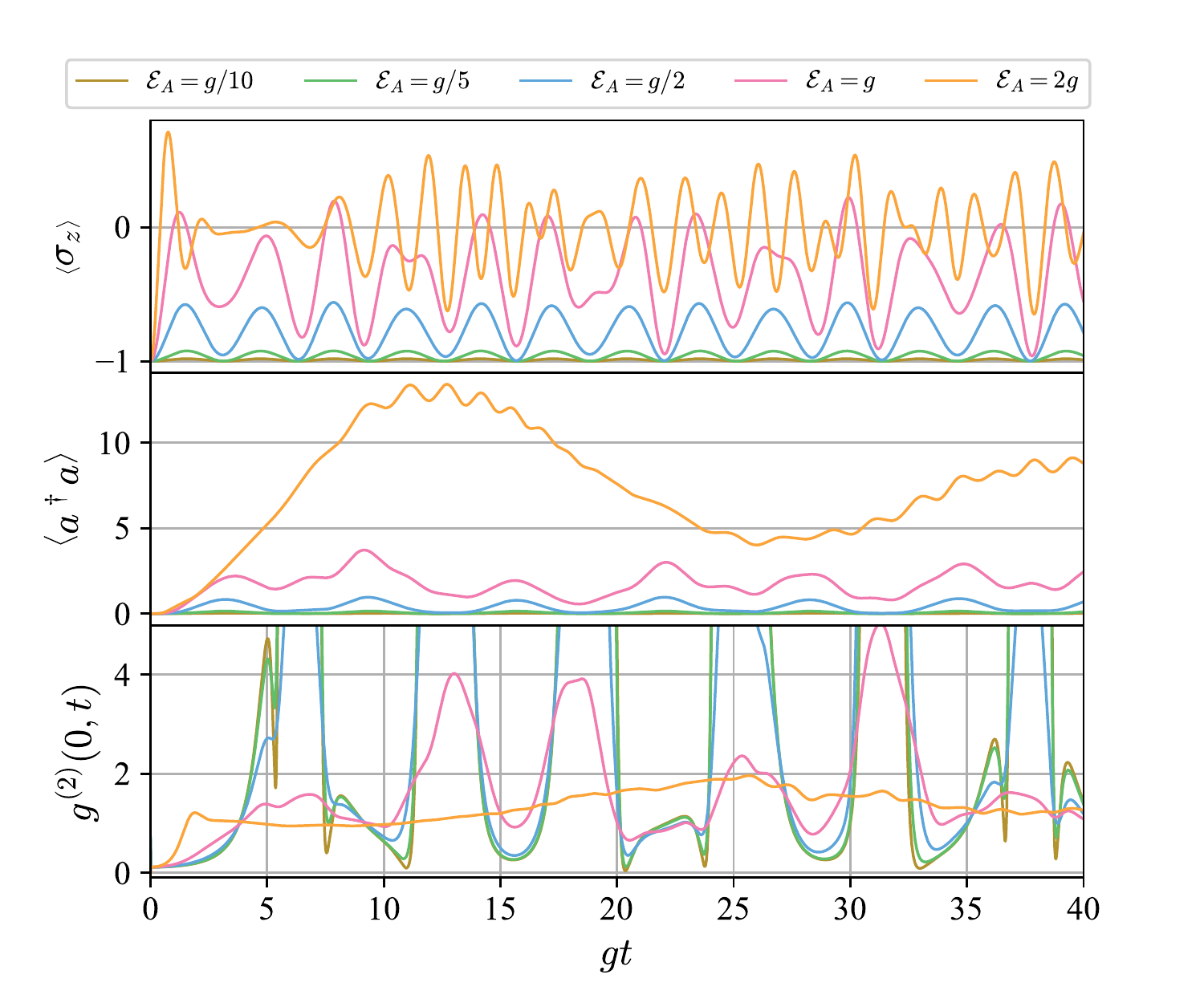}
\vspace{-.6cm}
\caption{The time evolution of the system populations and the instantaneous second-order correlation function within the cavity for a closed Jaynes-Cummings system for various TLS driving strengths.}
\label{fig:closed_atom_drive}
\vspace{.2cm}
\includegraphics[trim={0cm 0cm 0cm 1cm},clip,width=.7\textwidth]{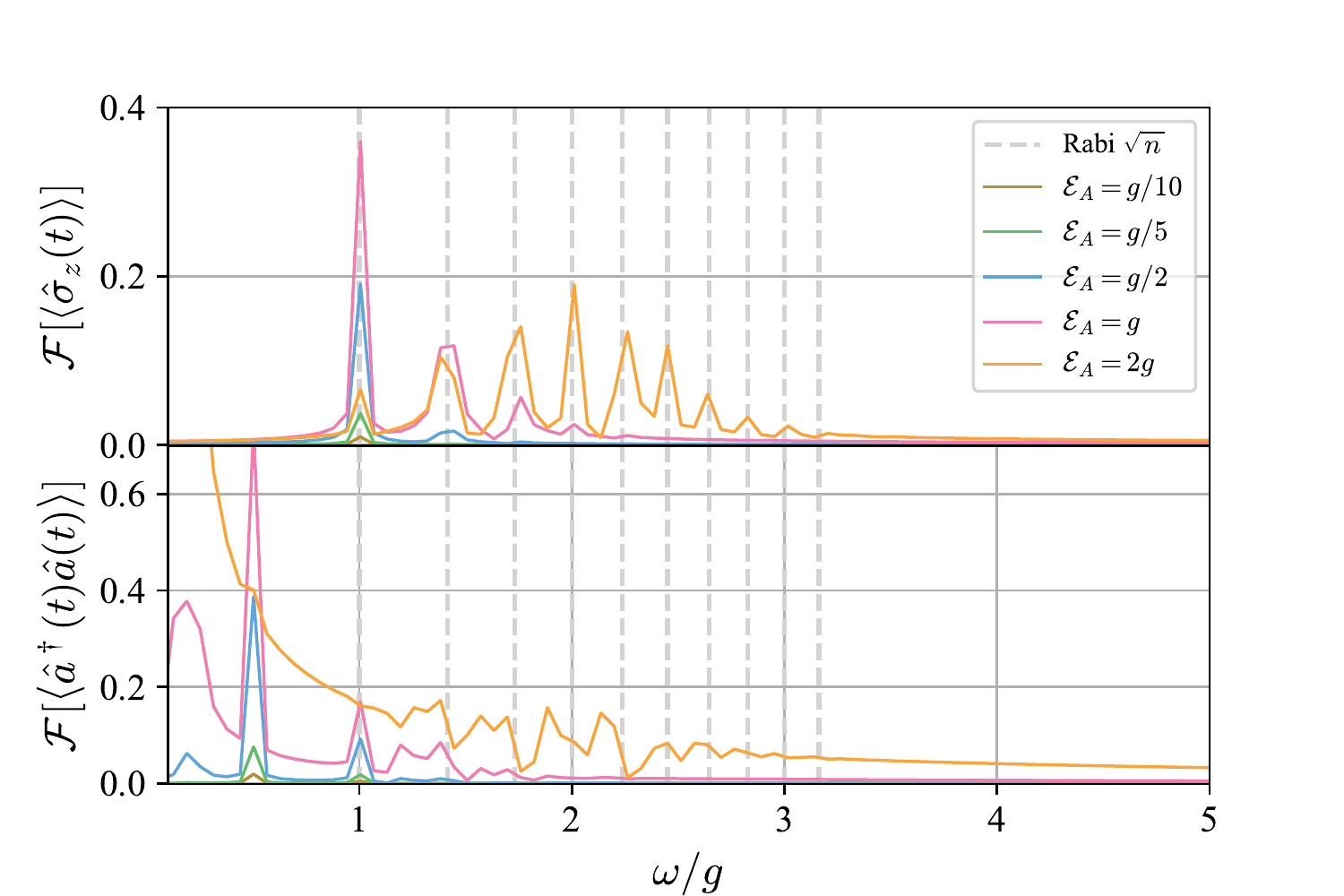}
\vspace{-.4cm}
\caption{The Fourier transforms of the time traces of the upper 2 panels in FIG. \ref{fig:closed_atom_drive}}
\label{fig:closed_atom_drive_FT}
\end{figure}

In order to understand the collapse-revival-type features presented in the main text in FIG. 5 we looked at the dynamics of the closed system at various driving strength as seen in FIG. \ref{fig:closed_atom_drive}. For small $\mathcal{E}_A\ll g$ the atomic population oscillates at the Rabi frequency, whereas the cavity population oscillates with half the frequency. As the driving strength is increased, more and more frequencies participate in the dynamics. At $\mathcal{E}_A>g$ the population inversion starts to show a revival after an initial collapse. Increasing the driving strength further makes these revivals cleaner, resembling more and more the case without driving.

An explanation to this behaviour can be found by looking at the eigenstates of the system with coherent TLS driving. These are displaced Rabi doublets where the amplitude of the coherent displacement is $\mathcal{E}_A/g$ \cite{Alsing1992Stark}. When this displacement is small, the system behaves closer to standard Rabi oscillations, whereas for very large driving strengths the coherent state characteristics dominate. This tendency can also be observed in the instantenous second-order correlation function for the cavity field which is defined as 
\begin{align}
    g^{(2)}(0,t)=\frac{\av{\lk\adop(t)\aop(t)\rk}^2}{\av{\adop(t)\aop(t)}^2}.
\end{align}
As the driving strength increases, this function tends to $1$, a characteristic of a coherent field.

The cavity field behaves differently from the TLS population as in this case the two superposition states for strong driving are closer to each other than without driving. Therefore, the revivals are observed only as perturbations on top of the main dynamics.

In order to show the above described features from a different perspective, we also calculate the Fourier transforms of the above time traces in FIG. \ref{fig:closed_atom_drive_FT}. Here, the TLS population shows more of the characteristic Rabi frequencies with increasing driving strengths resulting in the collapse-revival type dynamics. Meanwhile, the cavity field shows extra frequency contributions as well.

\clearpage
\subsection{Time evolution with constructive feedback}
\begin{figure}[h!]
\centering
\vspace{-.3cm}
\includegraphics[trim={0cm 0cm 0cm .7cm},clip,width=.7\textwidth]{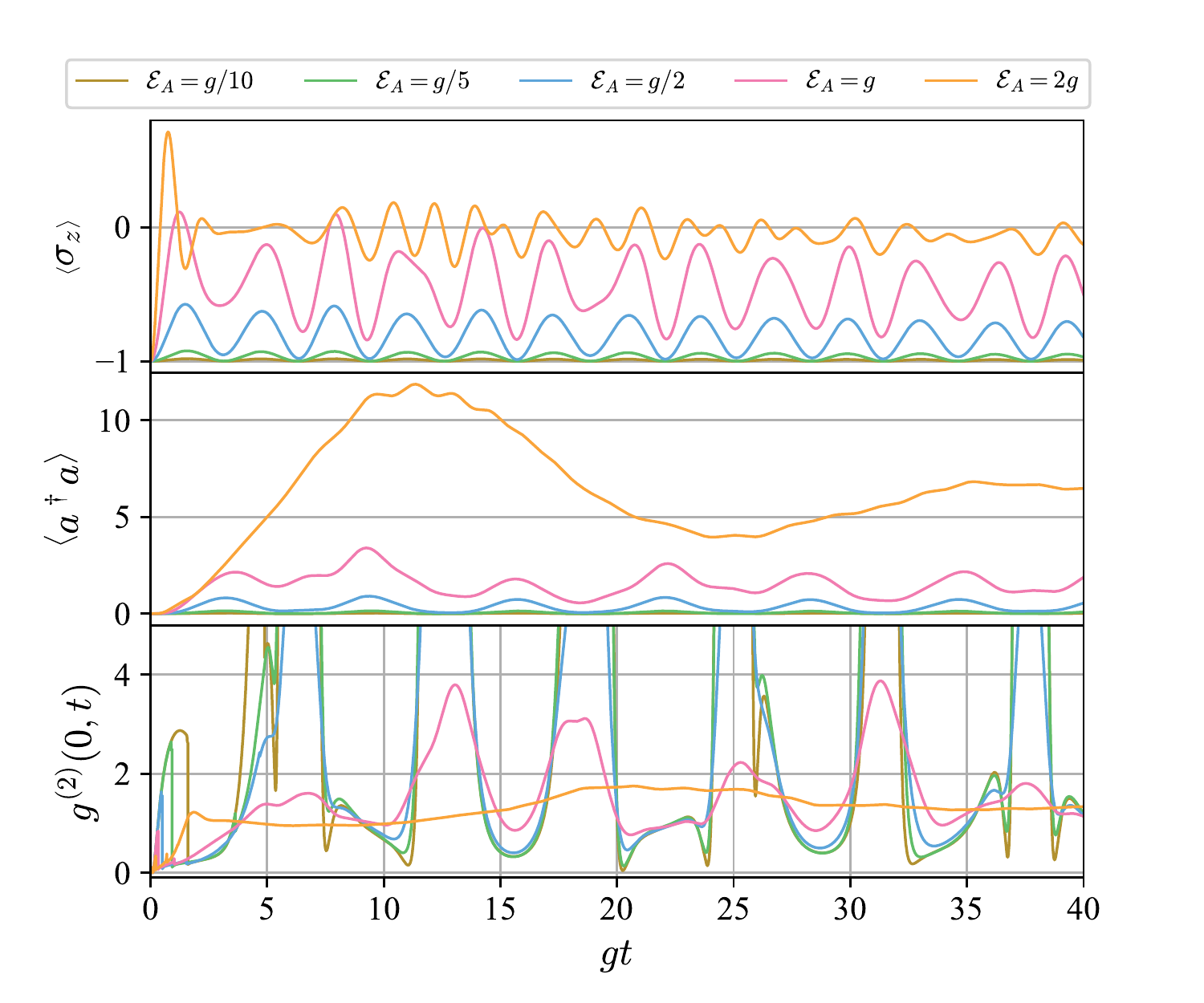}
\vspace{-.6cm}
\caption{The time evolution of the system populations and the instantaneous second-order correlation function within the cavity for a Jaynes-Cummings system with constructive feedback for various TLS driving strengths. Parameters: $\kappa_1=4\kappa_2=2g/25, g\tau=20$}
\label{fig:fb_atom_drive}
\vspace{-.3cm}
\end{figure}
\begin{figure}[h!]
\centering
%\vspace{-.3cm}
\includegraphics[trim={0cm 0cm 0cm 1cm},clip,width=.7\textwidth]{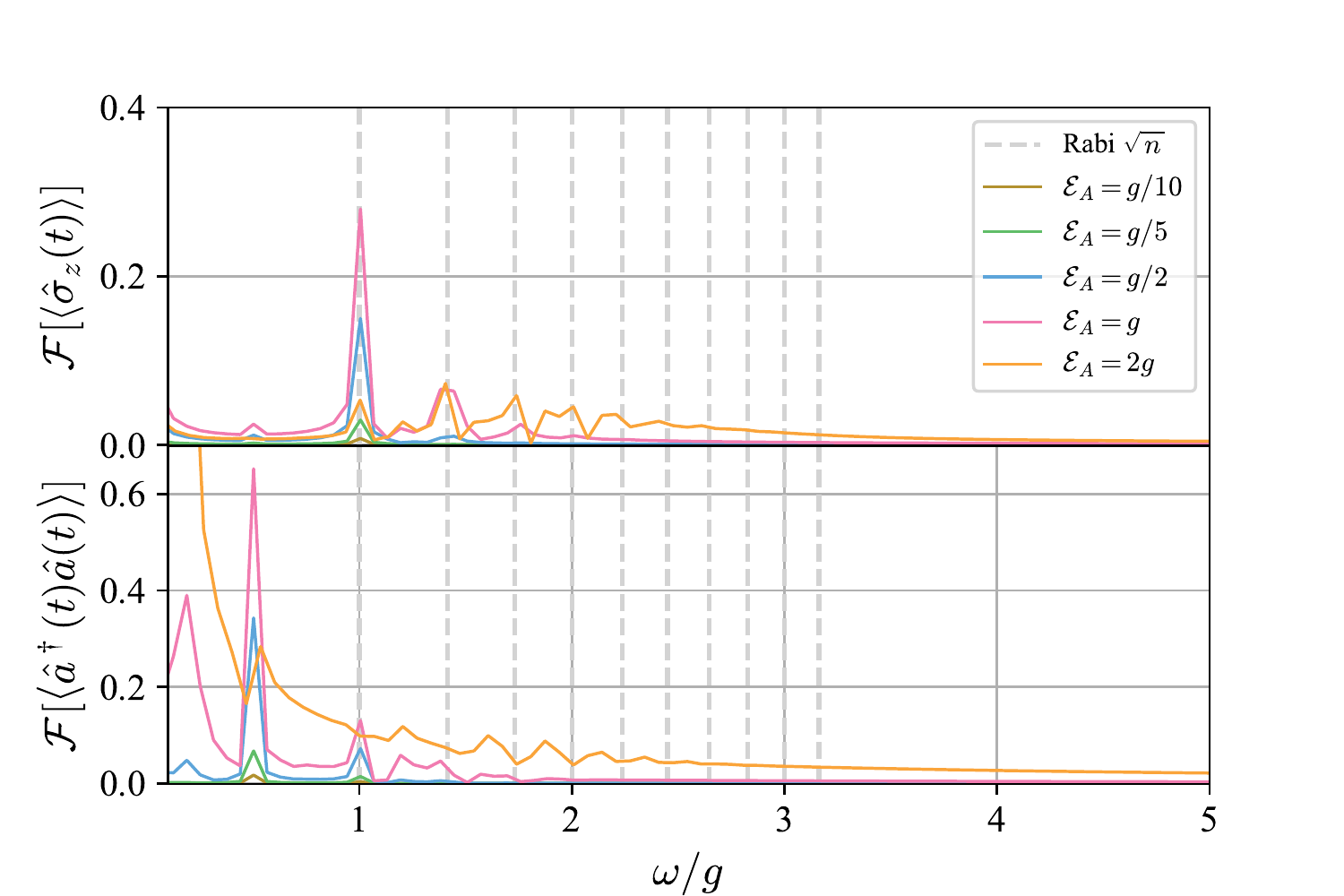}
\vspace{-.4cm}
\caption{The Fourier transforms of the time traces of the upper 2 panels in FIG. \ref{fig:fb_atom_drive}}
\label{fig:fb_atom_drive_FT}
\vspace{-.3cm}
\end{figure}

Let us compare the results of the previous subsection with the constructive feedback case. The timetraces in FIG. \ref{fig:fb_atom_drive} show similar features. The Fourier transform in FIG. \ref{fig:fb_atom_drive_FT}, however, reveals that the extra frequencies, present only for the cavity in the closed system case, appear in the TLS population dynamics as well for constructive feedback.

One reason for this difference is that beyond the coherent contributions of the driving field and the cavity, a delayed returning coherent contribution which originates from the cavity field also drives the TLS population.

\clearpage
\bibstyle{apsrev4-1}
\bibliography{citations}